\documentclass[conference]{IEEEtran}
\pdfoutput=1

\usepackage{xcolor, soul}
\usepackage{xspace}
\usepackage{multirow}
\usepackage{tabularx}
\usepackage{tikz}
\usepackage[bottom]{footmisc}
\newcommand*\circled[1]{\tikz[baseline=(char.base)]{\node[shape=circle,fill,inner sep=0.5pt] (char) {\textcolor{white}{#1}};}}
\usepackage[most]{tcolorbox}
\usepackage{amsmath}
\usepackage{enumitem}

\usepackage{subcaption}
\usepackage{fancyhdr}
\usepackage{mathtools}

\usepackage{booktabs}

\definecolor{darkspringgreen}{rgb}{0.09, 0.45, 0.27}
\definecolor{denim}{rgb}{0.08, 0.38, 0.74}
\definecolor{darkolivegreen}{rgb}{0.33, 0.42, 0.18}
\definecolor{tangerine}{rgb}{0.95, 0.52, 0.0}
\definecolor{mahogany}{rgb}{0.75, 0.25, 0.0}
\definecolor{coolblack}{rgb}{0.0, 0.18, 0.39}
\definecolor{darkpink}{rgb}{0.88, 0.28, 0.54}

\definecolor{seagreen}{rgb}{0.18, 0.55, 0.34}

\usepackage[colorinlistoftodos,prependcaption,textsize=scriptsize,textwidth=8mm]{todonotes} %
\usepackage{marginnote} 

\newcommand\revmid[1]{\todo[linecolor=magenta,backgroundcolor=magenta!15,bordercolor=magenta]{\textcolor{black}{\textbf{#1}}}}

\renewcommand\revmid[1]{}

\newcommand{\js}[1]{{\color{black}{#1}}} %

\definecolor{pred}{rgb}{0.7843, 0.0039, 0.3137} %

\definecolor{darkpink}{rgb}{0.88, 0.28, 0.54}
\definecolor{forestgreen}{rgb}{0.0, 0.27, 0.13}
\definecolor{amber}{rgb}{1.0, 0.49, 0.0}

\newcommand{\joel}[1]{{\color{cyan}#1}}

\newcommand{\inum}[1]{(\textit{#1})\xspace}

\newcommand{\sects}[1]{{§#1}\xspace} %
\newcommand{\sect}[1]{{§#1}\xspace} %
\newcommand{\head}[1]{{\noindent\textbf{#1.}\xspace}} %
\newcommand{\figs}[1]{{Figs.~#1}\xspace} %
\newcommand{\fig}[1]{{Fig.~#1}\xspace} %

\newcommand\proposal{MetaStore\xspace}

\newcommand\dram{\texttt{\omciv{No-I/O}}\xspace}

\newcolumntype{Y}{>{\centering\arraybackslash}X}
\usepackage{tikz}

\newcommand{\squishlist}{
 \begin{list}{$\circ$}
  { \setlength{\itemsep}{0pt}
     \setlength{\parsep}{0pt}
     \setlength{\topsep}{3pt}
     \setlength{\partopsep}{0pt}
     \setlength{\leftmargin}{1em}
     \setlength{\labelwidth}{1em}
     \setlength{\labelsep}{0.5em} } }

\newcommand{\squishend}{
  \end{list}  }

\usepackage[compact]{titlesec}
\titlespacing\section{0pt}{5pt plus 2pt minus 2pt}{0pt plus 2pt minus 2pt}
\titlespacing\subsection{0pt}{5pt plus 2pt minus 2pt}{0pt plus 2pt minus 2pt}
\titlespacing\subsubsection{0pt}{5pt plus 2pt minus 2pt}{0pt plus 2pt minus 2pt}
\makeatletter
\g@addto@macro{\normalsize}{%
  \setlength{\abovedisplayskip}{2pt plus 0.5pt minus 1pt}
  \setlength{\belowdisplayskip}{2pt plus 0.5pt minus 1pt}
  \setlength{\abovedisplayshortskip}{0pt}
  \setlength{\belowdisplayshortskip}{0pt}
  \setlength{\intextsep}{2pt plus 1pt minus 1pt}
  \setlength{\textfloatsep}{2pt plus 1pt minus 1pt}
  \setlength{\skip\footins}{2pt plus 1pt minus 1pt}}
\makeatother

 \usepackage{setspace}
\usepackage[colorinlistoftodos,prependcaption,textsize=scriptsize]{todonotes}
\usepackage{todonotes}

\usepackage{blindtext,graphicx}
\usepackage[absolute]{textpos}
\usepackage{datetime}

\definecolor{seagreen}{rgb}{0.18, 0.55, 0.34}
\definecolor{ballblue}{rgb}{0.13, 0.67, 0.8}

\newcommand\omciv[1]{{{#1}}}

\definecolor{darkgreen}{rgb}{0.0, 0.44, 0.34}
 
 \newcommand\ssdc{\texttt{SSD-C}\xspace}
 \newcommand\ssdp{\texttt{SSD-P}\xspace}
 
\sloppypar

\newcommand\unfinished[1]{{\color{black}{#1}}}

\newcommand\hm[1]{{\color{violet}{#1}}}

\newcommand\randomio{\textsf{R-Qry}\xspace}
\newcommand\streamio{\textsf{S-Qry}\xspace}

\definecolor{dollarbill}{rgb}{0.52, 0.73, 0.4}

\newcommand\cmashopt{\textsf{KSS}\xspace}

\renewcommand{\js}[1]{{\color{black}{#1}}} %
\renewcommand{\joel}[1]{{\color{black}#1}}
\renewcommand\hm[1]{{\color{black}{#1}}}

\definecolor{indiagreen}{rgb}{0.07, 0.53, 0.03}

\newcommand\rev[1]{{\color{black}{#1}}} %
\newcommand{\new}[1]{{\color{black}{#1}}}  %

\newcommand\revhid[1]{\todo[linecolor=magenta,backgroundcolor=magenta!15,bordercolor=magenta]{\textcolor{black}{\textbf{#1}}}}

\newcommand\gram[1]{{\color{black}{#1}}}

\renewcommand\revhid[1]{}

\usepackage[sort,compress]{cite}
\usepackage{amsmath,amssymb,amsfonts}
\usepackage{algorithmic}
\usepackage{graphicx}
\usepackage{textcomp}
\usepackage{xcolor}
\usepackage[hyphens]{url}
\usepackage{fancyhdr}
\usepackage{hyperref}

\def\BibTeX{{\rm B\kern-.05em{\sc i\kern-.025em b}\kern-.08em
    T\kern-.1667em\lower.7ex\hbox{E}\kern-.125emX}}

\title{\huge{\proposal: High-Performance Metagenomic Analysis\\via In-Storage Computing}}

\DeclareRobustCommand*{\IEEEauthorrefmark}[1]{%
  \raisebox{0pt}[0pt][0pt]{\textsuperscript{\footnotesize\ensuremath{#1}}}}
\newcommand{\affilETH}{\IEEEauthorrefmark{1}}
\newcommand{\affilPOSTECH}{\IEEEauthorrefmark{2}}

\author{
    \IEEEauthorblockN{%
    Nika Mansouri Ghiasi\affilETH \quad
    Mohammad Sadrosadati\affilETH \quad
    Harun Mustafa\affilETH \quad
    Arvid Gollwitzer\affilETH \quad
    }\vspace{0.1em}
    \IEEEauthorblockN{%
    Can Firtina\affilETH \quad
    Julien Eudine\affilETH \quad
    Haiyu Ma\affilETH \quad
    Jo\"{e}l Lindegger\affilETH \quad
    Meryem Banu Cavlak\affilETH \quad
    }\vspace{0.1em}
    \IEEEauthorblockN{
    Mohammed Alser\affilETH \quad
    Jisung Park\affilPOSTECH \quad
    Onur Mutlu\affilETH
    }\\
    
     \IEEEauthorblockA{\IEEEauthorrefmark{1}ETH Zürich\hspace{0.5em} \IEEEauthorrefmark{2}POSTECH}
 }

\begin{document}
\bstctlcite{IEEEexample:BSTcontrol}
\maketitle
\thispagestyle{plain}
\pagestyle{plain}

\begin{abstract}
Metagenomics, the study of the genome sequences of diverse organisms in a common environment, 
has led to significant advancements in many 
fields.  
Since
the \hm{species} present in \hm{a} metagenomic sample are not known in advance, metagenomic analysis commonly involves the key tasks of determining the \hm{species} present in a sample and their relative abundances.  
These tasks require searching large metagenomic databases
containing information on different \hm{species}’ genomes. 
Metagenomic analysis suffers from significant data movement
overhead due to moving large amounts of low-reuse data from the storage system to the rest of the system.
In-storage processing 
can be a fundamental solution for reducing data movement overhead. However, designing an in-storage processing system for metagenomics is challenging because none of the existing approaches can be directly implemented in \new{storage}
effectively due to the hardware limitations of modern SSDs.

We propose \textbf{\proposal}, the \emph{first} in-storage processing system designed to significantly reduce the data movement overhead of end-to-end metagenomic analysis. \proposal is enabled by our lightweight and cooperative design that effectively leverages and orchestrates processing inside and outside the storage system.
Through our detailed analysis of the end-to-end metagenomic analysis pipeline and careful hardware/software co-design, we address in-storage processing challenges for metagenomics 
via specialized and efficient 
1)~task partitioning,
2)~data/computation flow coordination,
3)~storage technology-aware algorithmic optimizations,
4)~light-weight in-storage accelerators, and
5)~data mapping. 
Our evaluation shows that \proposal outperforms the state-of-the-art performance- and accuracy-optimized software metagenomic tools
by 2.7--\new{37.2}$\times$ and 6.9--\new{100.2}$\times$, respectively,  while matching the accuracy of the accuracy-optimized tool. \proposal achieves 1.5--5.1$\times$ speedup compared to the state-of-the-art metagenomic hardware-accelerated tool, while achieving significantly higher accuracy.

\end{abstract}

\section{Introduction}
\label{sec:introduction}

Metagenomics is an increasingly important domain in bioinformatics, which analyzes the genome sequences of various organisms of \emph{different species} present in a common environment (e.g., human gut, soil, or oceans)~\cite{hhrlich2011metahit,sunagawa2015structure,fierer2017embracing}. 
\js{Unlike} traditional genomics~\cite{cali2020genasm,casa23micro,li2018minimap2,ham2020genesis,turakhia2018darwin,gupta2019rapid} \js{that} studies genome sequences \js{from} an individual \js{(}or a small group of individuals\js{)} of the \emph{same species} \js{using a \emph{known} reference genome,} metagenomics deals with \js{genome sequences whose species are \emph{not known} in advance in many cases, thereby requiring comparisons of} 
the \js{target} sequences against large databases of many reference genomes. 
Metagenomics has led to groundbreaking advancements in many fields, such as precision medicine~\cite{kintz2017introducing,dixon2020metagenomics}, urgent clinical settings~\cite{taxt2020rapid}, understanding microbial diversity of an environment~\cite{afshinnekoo2015geospatial,hsu2016urban}, discovering early warnings of communicable diseases~\cite{john2021next,nagy2021targeted,nieuwenhuijse2017metagenomic}, and outbreak tracing~\cite{hadfield2018nextstrain}. 
The pivotal role of metagenomics, together with the rapid \js{improvements} in genome sequencing technologies (e.g., reduced costs and improved throughput~\cite{berger2023navigating}), has resulted in a fast-growing adoption of metagenomics over the past few years~\cite{dixon2020metagenomics,chiang2019from,chiu2019clinical}.

The metagenomic workflow consists of three key steps\js{: \inum{i}~sequencing, \inum{ii}~basecalling, and \inum{iii} metagenomic analysis.} 
First, \js{sequencing extracts} the genomic information of all organisms in a sample into digital data. 
\js{S}ince \js{the} current sequencing \js{technologies \emph{cannot} process} the DNA molecule as a whole, \js{a sequencing machine} generate\js{s} randomly sampled, inexact fragments \js{of genomic information, called \emph{reads}}. 
Second, basecalling converts the raw sequencer data of \js{reads} into \js{a} sequence of characters that represent nucleotides\js{, \texttt{A}, \texttt{C}, \texttt{G}, and \texttt{T}}. 
\js{T}hird, metagenomic analysis determines the species \emph{present/absent} in the sample and their \emph{relative abundances}.

To enable fast and efficient metagenomics for many critical applications, it is essential to improve the performance and energy efficiency of metagenomic analysis due to three reasons. 
First, metagenomic analysis is typically performed much more frequently compared to the other two steps \js{ (i.e., sequencing and basecalling)} in the metagenomic workflow.
\js{While the two steps} are one-time tasks for a sample in many cases\js{, sequenced and basecalled reads in a sample} can often be analyzed \js{in} \emph{multiple studies} or \js{at} \emph{different times} in the same study~\cite{bokulich2020measuring}. 
Second, as shown in our motivational analysis in \sect{\ref{sec:motivation}} on a high-end server node, even for one round of analysis, the analysis step bottlenecks the end-to-end performance and energy consumption of the workflow. 
Third, \js{the performance and energy-efficiency gaps between the analysis step and the other steps are expected to widen even more due to the rapid advancements in sequencing technologies, such as}
significant increases in throughput and energy efficiency of sequencing at rates~\cite{hu2021next,alser2020accelerating,katz2021sra,enastats,alser2022molecules,leinonen2010sequence} higher than Moore's Law~\cite{berger2023navigating}, the development of sequencing technologies that enable analysis during sequencing~\cite{zhang2021real,firtina2023rawhash,kovaka2020targeted, mutlu2023accelerating,firtina2023rawhash2,lindegger2023rawalign}, and portable and energy-efficient high-throughput sequencers~\cite{minion21,jain2016oxford}. 
Due to these reasons, it is not feasible to efficiently keep up with these advancements by merely scaling up traditional systems to improve the analysis.

The analysis step suffers from significant data movement overhead due to accessing large amounts of low-reuse data. Since we do not know the species present in a metagenomic sample, metagenomic tasks require searching large databases (up to several terabytes~\cite{ncbi2023,karasikov2020metagraph,shiryev2023indexing,pebblescout}, with emerging databases exceeding a hundred terabytes~\cite{shiryev2023indexing,pebblescout}) that contain information on different organisms' reference genomes. Based on the recent trends, the database sizes are expected to increase even further in the future, and at a fast pace, due to recent advances in the automated and scalable construction of genomic data from more organisms~\cite{rautiainen2023telomere,jarvis2022semi}.\footnote{For example, based on recently published trends, the ENA assembled/annotated sequence database currently \emph{doubles} every 19.9 months~\cite{enastats}, and the BLAST nt database doubled from 2021 to 2022~\cite{ntdouble}.} Our motivational analysis (\sect{\ref{sec:motivation}}) of the state-of-the-art metagenomic analysis tools shows that data movement overhead from the storage system significantly impacts their end-to-end performance, even when the main memory is larger than the accessed data. Due to its low reuse, the data needs to move all the way from the storage system to the main memory and processing units for its first use, and most likely not be used again within the analysis. 
This low data reuse, alongside low computation intensity and limited I/O bandwidth leads to large storage I/O overhead.

While there have been great efforts in accelerating metagenomic analysis, to our knowledge, no prior work fundamentally addresses its storage I/O overheads. Some works~\cite{kim2016centrifuge,wood2019improved,pockrandt2022metagenomic,muller2017metacache} aim to alleviate this overhead by applying sampling techniques to reduce the database size, but they incur accuracy loss, which is problematic for many use cases~\cite{milanese2019microbial,salzberg2016next,gihawi2023major,pockrandt2022metagenomic}. 
Various other works~\cite{dashcam23micro, kobus2021metacache,wang2023gpmeta,wu2021sieve, hanhan2022edam,kobus2017accelerating,shahroodi2022krakenonmem,shahroodi2022demeter,armstrong2022swapping,jia2011metabing} accelerate other bottlenecks in metagenomic analysis, such as computation and main memory bottlenecks. These works do not alleviate I/O overhead, whose impact on end-to-end performance (as shown in \sect{\ref{sec:motivation}}) becomes even larger when other bottlenecks get alleviated.

\js{\emph{In-storage processing (ISP)}, i.e., p}rocessing data directly in\js{side} the storage \js{device} where \js{target} data resides\js{,} can be a fundamental and cost-efficient approach for \js{mitigating} the \js{data-movement bottleneck in metagenomic analysis, given its} three key benefits. 
First, \js{ISP can significantly} reduce unnecessary data movement from storage system\js{s by processing large amounts of low-reuse data inside the storage while sending only the results to the host}. 
Second, \js{ISP} can \js{leverage} the \js{high SSD-}internal bandwidth \js{to access target data} without being restricted by the SSD-external bandwidth limits. 
Third, it reduces the overall burden of the application from the rest of the system (e.g., processing units and main memory), freeing up the host system to perform other useful work meanwhile.

Despite \js{the} benefits \js{of ISP}, none of the existing approaches \js{for metagenomic analysis} can be directly implemented as an ISP system effectively due to \js{the limited} hardware resources available in storage \js{devices}.
Some tools incur a large number of random accesses to search the database~\cite{wu2021sieve,wood2019improved}, which hinders leveraging ISP's large potential by preventing the full utilization of the SSD's\footnote{In this work, we focus on NAND flash-based SSDs, the predominant technology used in the modern storage systems. We expect that our insights and our designs would also provide benefits with storage systems built with other emerging technologies.} internal bandwidth. This is due to costly conflicts in internal SSD resources (e.g., channels and NAND flash chips~\cite{nadig2023venice,tavakkol2018flin,kim2022networked}) caused by random accesses. 
Some tools predominantly incur more suitable streaming accesses~\cite{lapierre2020metalign,pockrandt2022metagenomic}, 
but they do so at the cost of more computation and main memory capacity requirements. 
This becomes challenging for ISP due to limited hardware resources available inside SSDs.
Therefore, directly adopting these approaches in ISP incurs performance, energy, and storage device lifetime overheads.

Our goal is to improve metagenomics performance by reducing data movement overhead, right from the storage system, in a cost-effective manner and by addressing the challenges of ISP. To this end, we propose \textbf{\proposal}, the \emph{first} ISP system designed for metagenomics. 
The key idea of \proposal is to enable  \emph{cooperative} ISP for metagenomics, where we do not merely focus on the storage system and, instead, capitalize on the strengths of processing both inside and outside the storage system via an end-to-end and synergistic design.

We design \proposal as an efficient pipeline between the SSD and the host system to \inum{i}~\emph{leverage} and \inum{ii}~\emph{orchestrate} the capabilities of both. To this end, we propose new hardware/software co-designed approaches in five aspects.
First, based on our detailed analysis of the key metagenomic tasks,
we partition and map the task parts to the host and the ISP system, such that each part maps to its most suitable architecture. 
Second, we coordinate the data/computation flow between the host and the SSD, such that \proposal \inum{i}~completely overlaps the data transfer time between them with computation time to reduce the communication overhead between the task partitions, \inum{ii}~can leverage both the SSD's internal and external bandwidth efficiently, \inum{iii}~does not require large DRAM inside the SSD or a large number of writes to the flash chips. 
Third, we devise storage technology-aware algorithm optimizations to enable efficient access patterns to the SSD. 
Fourth, we design lightweight in-storage accelerators to perform \proposal's ISP tasks, while minimizing the required SRAM/DRAM buffer spaces inside the SSD. 
Fifth, we design an efficient data mapping scheme and Flash Translation Layer (FTL), specialized based on the characteristics of the metagenomic tasks to leverage the SSD's full internal bandwidth.

We evaluate \proposal with two different SSD configurations (performance-optimized~\cite{samsungPM1735} and cost-optimized~\cite{samsung870evo}) to show the impact of the storage system.
We compare \proposal against three state-of-the-art metagenomic analysis tools: \inum{i}~Kraken2~\cite{wood2019improved}, which is optimized for performance, \inum{ii}~Metalign~\cite{lapierre2020metalign}, which is optimized for accuracy, and 
\inum{iii}~a state-of-the-art processing-in-memory accelerator, Sieve~\cite{wu2021sieve}, integrated into Kraken2 to accelerate its k-mer matching. By analyzing \emph{end-to-end} performance, we show that
\proposal provides 2.66--37.24$\times$ speedup compared to Kraken2, and 1.46--5.06$\times$ speedup compared to Sieve, while providing significantly higher accuracy.
\proposal provides  6.93--100.20$\times$ speedup compared to Metalign, while providing the same accuracy. 
\proposal provides large average energy reduction
of 5.36$\times$ and 1.88$\times$ compared to Kraken2 and Sieve, respectively, and 15.16$\times$ compared to accuracy-optimized Metalign. \proposal's benefits come at a low area cost of 1.7\% of the three cores~\cite{cortexr4} in a SSD controller~\cite{samsung860pro}. 

This work makes the following major contributions:
\begin{itemize}[leftmargin=*, noitemsep, topsep=0pt]
    \item We propose \proposal, the \emph{first} ISP system for metagenomics. Via our end-to-end and cooperative approach, \proposal fundamentally reduces I/O costs of metagenomic analysis. 

    \item We present new hardware/software co-designed approaches to enable an efficient pipeline between the host and the SSD to address the challenges of ISP for metagenomics.
    \item \proposal improves the performance and energy efficiency compared to the state-of-the-art metagenomics tools, while maintaining high accuracy. It does so without relying on costly hardware resources throughout the sytsem, making metagenomics more accessible for its wider adoption. 
\end{itemize}

\section{Background}
\label{sec:background}

\subsection{Metagenomics Analysis}
\label{sec:background-metagenomics}

Metagenomic analysis tools commonly perform two key tasks of identifying \inum{i}~species \emph{present/absent} in a sample, and
\inum{ii} their \emph{relative abundances}.

\subsubsection{Presence/absence Identification} 

To find species present in the sample, many tools extract \emph{k-mers} (i.e., subsequences of length $k$) from the input query reads and search for the k-mers in a \emph{reference database}. 
Each database contains k-mers extracted from reference genomes of a wide range of species. The database \rev{associates} each indexed k-mer with a \emph{taxonomic identifier (i.e., tax ID)}\footnote{A tax ID is an integer attributed to a cluster of related species.} of the reference genome(s) the k-mer comes from. 
The GB- or TB-scale databases typically support  random (e.g.,~\cite{wood2019improved,kim2016centrifuge,milanese2019microbial}) or streaming (e.g.,~\cite{lapierre2020metalign,pockrandt2022metagenomic,lemane2022kmtricks}) access patterns. 

\head{Tools with Random Access Queries (\randomio)}
Many tools commonly perform random accesses to search their database.
A state-of-the-art tool in this category is Kraken2~\cite{wood2019improved}, which maintains a hash table associating each indexed k-mer to a tax ID.  To identify which species are present in a set of queries, Kraken2 extracts k-mers from the read queries and searches the database to retrieve the k-mers' associated tax IDs. 
For each read, Kraken2 collects the tax IDs of that read's k-mers and, based on the occurrence frequencies of these tax IDs,  uses a classification algorithm to assign a single tax ID to each read. Finally, Kraken2 identifies the species with these tax IDs as present in the sample.

\head{Tools with Streaming Access Queries (\streamio)}
Some tools (e.g., ~\cite{lapierre2020metalign,pockrandt2022metagenomic,lemane2022kmtricks}) predominantly feature streaming accesses to their databases. A state-of-the-art tool in this category is Metalign~\cite{lapierre2020metalign}. Presence/absence identification in Metalign is done via 1) preparing the input read set queries, and 2) finding species present in them. To process the read set queries, the tool extracts k-mers from the reads and sorts them. Finding the species present in the sample involves two steps.  First, it finds the intersecting k-mers between the query k-mers and a pre-sorted reference database. In this step, the tool uses large k-mers (e.g., $k=60$) for both the queries and the database to maintain a low species false positive rate. This is because large k-mers are more unique, and matching a long k-mer ensures that the queries have at least one long and specific match to the database. Second, the tool finds the tax IDs of the intersecting k-mers. To do so, it searches for the intersecting k-mers or \hm{their} prefixes in a smaller \emph{sketch database} of variable-sized k-mers. Each \emph{sketch} is a small representative subset of k-mers associated with a given tax ID. Searching for prefixes in this step increases the true positive rate by expanding the number of matching tax IDs.

\subsubsection{Abundance Estimation}

After finding the species present in the sample, some applications require a more sensitive step to find the species' relative abundances~\cite{lu2017bracken,sun2021challenges}. Different tools implement their own approaches for estimating abundances, ranging from relatively light-weight statistical models~\cite{lu2017bracken,dimopoulos2022haystac} to more accurate but computationally-intensive read mapping~\cite{lapierre2020metalign,kim2016centrifuge,milanese2019microbial}. 
Read mapping is the process of finding potential matching locations of reads against one or more reference genomes. Metagenomic tools can map the query reads against reference genomes of species present in the sample,
accurately determining the number of reads belonging to each species.

\subsection{SSD Organization}
\label{sec:background-ssd}

\begin{figure}[b]
         \centering
         \includegraphics[width=0.8\columnwidth]{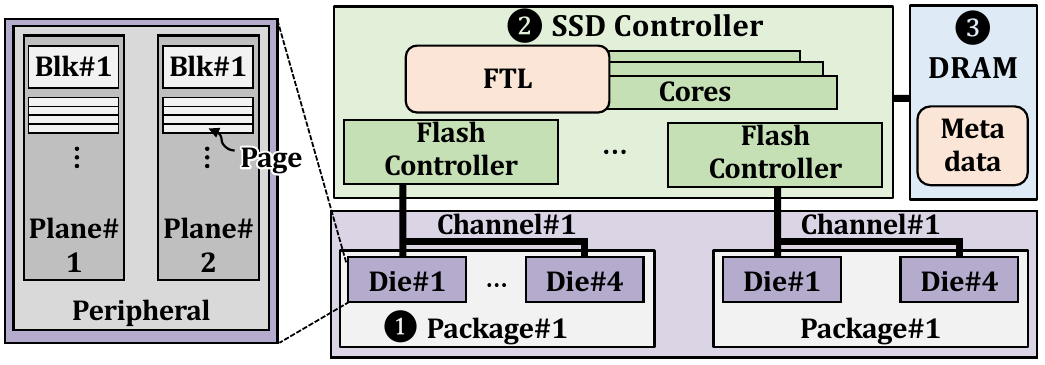}
         \caption{Organizational overview of a modern SSD.}
         \label{fig:ssd}
\end{figure} 

\fig{\ref{fig:ssd}} depicts the organization of a modern NAND flash-based SSD, consisting of three main components.

\head{\circled{1}~NAND Flash Memory}
A NAND package consists of multiple \emph{dies} or \emph{chips}, sharing the NAND package's I/O pins. 
One or multiple packages share a command/data bus or \emph{channel} to communicate with the SSD controller.
Dies can operate independently, but each channel can be used by only one die at a time to communicate with the controller.
Each die has multiple (e.g., 2 or 4) \emph{planes}, each with thousands of \emph{blocks}.
Each block has hundreds to thousands of 4--16 KiB \emph{pages}.
NAND flash memory performs read/write operations at page granularity but erase operations at block granularity.
The peripheral circuitry to access pages is shared among the planes in each die. Hence, it is possible for the planes in a die to operate concurrently when accessing pages (or blocks) at the same offset. This mode is called the \emph{multiplane} operation.

\head{\circled{2}~SSD Controller}
An SSD controller consists of two key components. First, multiple cores run SSD firmware, commonly referred to as the \emph{flash translation layer (FTL)}. The FTL is responsible for communication with the host system, internal I/O scheduling, and various SSD management tasks. Second, per-channel hardware flash controllers manage request handling and error correction for the NAND flash chips.

\head{\circled{3}~DRAM} Modern SSDs use low-power DRAM to store metadata for SSD management tasks. Most of the \hm{DRAM} capacity inside the SSD (i.e., internal DRAM) is used to store the logical-to-physical (L2P) mappings, which are typically maintained at a granularity of 4KiB to enhance random access performance. Therefore, in a 32-bit architecture, and 4 bytes of metadata stored for every 4KiB of data, L2P memory overhead is about 0.1\% of the SSD's capacity. For example, a 4-GB LPDDR4 DRAM is used for a 4-TB SSD~\cite{samsung860pro}.

\section{Motivational Analysis}
\label{sec:motivation}

\subsection{Criticality of Metagenomic Analysis}

The analysis step bottlenecks the end-to-end performance of the metagenomic workflow and poses a pressing need for acceleration~\cite{wu2021sieve,hanhan2022edam,chiang2019from,edgar2022petabase,taxt2020rapid,sereika2022oxford} for three reasons. First, the sequencing and basecalling steps for a sample read set are one-time tasks in the majority of cases. In contrast, in many cases, the reads from a single sequenced sample can be analyzed by \emph{multiple studies} or \emph{at different times} in the same study~\cite{bokulich2020measuring}.
Second, even for one round of analysis, the analysis \emph{throughput} is significantly lower than the throughput with which modern sequencing machines generate data~\cite{illuminax,ontpromethion}. A single sequencing machine can sequence \emph{many samples} from different sources (e.g., different patients or environments) in parallel~\cite{hu2021next,shokralla2015massively}, achieving very high throughput.  Our analysis with a state-of-the-art metagenomic analysis tool~\cite{lapierre2020metalign} shows that analyzing the data sequenced and basecalled by a high-throughput sequencer in 48 hours takes 38 days on a high-end server node (details in \sect{\ref{sec:methodology}}). Such long analysis poses serious challenges for metagenomic analysis, specifically for the time-critical use cases (e.g., clinical settings~\cite{taxt2020rapid} and timely and widespread surveillance of infectious diseases~\cite{hadfield2018nextstrain}). Since the growth rate of sequencing throughput is higher than Moore's Law~\cite{berger2023navigating},
this already large gap between sequencing and computation throughput is widening~\cite{hu2021next,katz2021sra,enastats,leinonen2010sequence}, and merely scaling up traditional systems for analysis is not efficient. 
Third, the urgent need for rapid metagenomics has driven significant advancements in sequencing technologies that enable \emph{runtime analysis during sequencing}~\cite{zhang2021real,firtina2023rawhash,kovaka2020targeted, mutlu2023accelerating}, which increasingly highlights the criticality of fast analysis.

Similarly, the analysis step is the primary energy bottleneck in the metagenomic workflow, and optimizing its efficiency is vital as sequencing technologies rapidly evolve. For example, a high-end sequencer~\cite{illuminax} uses 405 KJ to sequence and basecall 100 million reads, with 92.5 Mbp/s throughput and 2,500 W power consumption~\cite{illuminax}. In contrast, processing this dataset on a commodity server (\sect{\ref{sec:methodology}}) requires 675 KJ, accounting for 63\% of total energy. The need to enhance the analysis' energy efficiency is further increasing for two reasons.
First, sequencing efficiency has improved markedly. For example, a new version of Illumina sequencer from 2023~\cite{illuminax} provides 44$\times$ higher throughput at only 1.5$\times$ larger power consumption compared to an older version~\cite{illumina} from 2020, resulting in much better sequencing energy efficiency. Therefore, solely relying on scaling up commodity systems to improve the analysis throughput worsens the analysis energy bottleneck significantly. Second, the increased adoption of compact \emph{portable DNA sequencers}~\cite{jain2016oxford} for on-site metagenomic analysis (e.g., in remote locations~\cite{pomerantz2018real} or for personalized bedside care~\cite{chiang2019from}) offers high-throughput sequencing with low energy costs. This trend further amplifies the need for energy- and cost-effective analysis that can match the portability and convenience of these portable yet powerful devices.

\subsection{Data Movement Overheads}
\label{sec:motivation-ovhd}

We conduct experimental analysis to assess the storage system's impact on the performance of metagenomics analysis.

 \head{Tools and Datasets} We analyze two state-of-the-art  tools 
 for performing the presence/absence identification: 
1)~Kraken2~\cite{wood2019improved}, which queries the large database with random access patterns (\randomio),
\footnote{We experiment with both techniques of accessing the database devised in the \randomio baseline~\cite{wood2014kraken} and report the best timing. The first technique uses mmap to access the database, while the second technique loads the entire database from the SSD to DRAM as the first step when the analysis starts. In this experiment, the second approach performs slightly better since, when analyzing our read set, the application requires accessing most parts of the database during its execution.}
and 2)~Metalign~\cite{lapierre2020metalign},
which mostly exhibits sequential streaming accesses to the database (\streamio). We use the best-performing thread count in each case.
We use a query sample with 100 million reads\footnote{This is in the upper quartile of metagenomic read set sizes~\cite{katz2021sra}.} (\texttt{RL\_S001\_\_insert\_270.fq}) from the CAMI dataset~\cite{meyer2021critical}, commonly used for profiling metagenomic tools. 
We generate a database based on \hm{microbial genomes drawn from NCBI's databases}~\cite{ncbi2020,lapierre2020metalign} using default parameters for each tool. For Kraken2~\cite{wood2019improved}, this results in a 293~GB database. For Metalign~\cite{lapierre2020metalign}, this results in a 701~GB k-mer database
and \gram{a} 6.9~GB sketch tree. To show the impact of recent trends in the rapid increase of database sizes, we also analyze larger k-mer databases (0.6~TB and 1.4~TB for Kraken and Metalign, respectively) that include genomes of more species.

\head{System Configurations} 
We use a high-end server with AMD EPYC 7742 CPU~\cite{amdepyc} and 1.5-TB DDR4 DRAM. Note that the DRAM size is larger than the size of all data accessed during the analysis by each tool and on each input dataset. This way, we can analyze the fundamental I/O data movement overhead of moving large amounts of low-reuse data from the storage system to the main memory without being limited by DRAM capacity.
We evaluate the impact of I/O overhead 
with:
1)~a cost-optimized SSD (\ssdc)~\cite{samsung870evo} with a SATA3 interface~\cite{SATA}, 
2)~a performance-optimized SSD (\ssdp)~\cite{samsungPM1735} with a PCIe Gen4 interface~\cite{PCIE4}, and
3)~a hypothetical configuration with zero performance overhead due to storage I/O (\dram). 
\ssdp provides an order-of-magnitude higher sequential-read bandwidth compared to \ssdc. 
However, scaling up storage capacity only using performance-optimized SSDs is challenging due to their much higher prices and fewer PCIe slots compared to SATA on servers.

\begin{figure}[b]
    \centering
    \includegraphics[width=\linewidth]{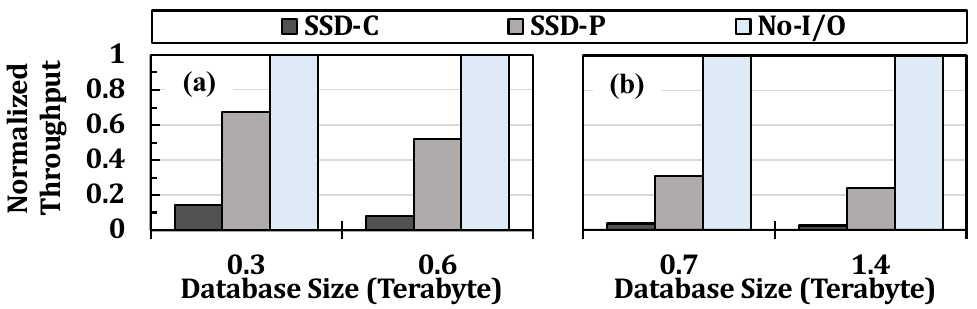}
    \caption{Performance of (a) \randomio and (b) \streamio under different storage configurations and database sizes.}
    \label{fig:motivation}
\end{figure}

\head{Results and Analysis}
\fig{\ref{fig:motivation}} shows the performance of \randomio and \streamio, normalized against the \dram configuration. We make three key observations. First, I/O overhead has a large impact on end-to-end performance for all cases. Compared to \ssdc (\ssdp), \dram leads to 9.4$\times$ (1.7$\times$) and 32.9$\times$ (3.6$\times$) better performance in \randomio and \streamio (on average across both databases), respectively.   
While both baselines significantly suffer from large I/O overhead, we observe a relatively larger impact on \streamio due to its lower data reuse compared to \randomio. 
Second, even using the state-of-the-art SSD (\ssdp) does not alleviate this overhead, leaving large performance gaps between \ssdp and \dram in both \randomio and \streamio.    
Third, I/O overhead increases as the databases grow larger. 
For example, for \randomio, the performance difference between \ssdc and \dram increases from 7.1$\times$ to 12.5$\times$ when the database grows from 0.3~TB to 0.6~TB. 
Based on these observations, we conclude that I/O accesses lead to large overheads in metagenomic analysis, with this overhead expected to worsen in the future.

This I/O overhead, stemming from the need to move large amounts of low-reuse data is a fundamental problem and hard to avoid. One might think it is possible to avoid I/O performance overheads by either 1) using sampling to shrink databases~\cite{kim2016centrifuge,wood2019improved,pockrandt2022metagenomic,muller2017metacache} or 2) keeping all data required by metagenomic analysis perpetually resident in main memory. Neither of these solutions is suitable. The first approach inevitably reduces accuracy~\cite{milanese2019microbial,meyer2021critical} to levels unacceptable for many 
use cases~\cite{milanese2019microbial,salzberg2016next,gihawi2023major,pockrandt2022metagenomic,berger2023navigating}.
The second approach is energy inefficient, costly, unscalable, and unsustainable due to two reasons. First, the sizes of metagenomic databases (which are already large, i.e., in some recent examples, exceeding a hundred terabytes~\cite{shiryev2023indexing,pebblescout}) have been increasing rapidly.
For example, recent trends show the \emph{doubling} of different important databases over the course of only several months~\cite{enastats,ntdouble}. Second, 
regardless of individual data sizes, different analyses need different databases. For example, a medical center may need to use various different databases for its patients (e.g., for different viral infections~\cite{centrifuge_db,kim2016centrifuge}, sepsis~\cite{taxt2020rapid}, etc.).
Therefore, it is inefficient and unsustainable to maintain \emph{all} data required by \emph{all possible} analyses in DRAM at all times.\footnote{Ultimately, these are the same reasons that the metagenomic community has been investigating storage efficiency (e.g., the mentioned sampling techniques~\cite{kim2016centrifuge,wood2019improved,pockrandt2022metagenomic,muller2017metacache}) as opposed to merely relying on scaling the system's DRAM.}

The impact of I/O accesses on end-to-end performance becomes even more prominent in emerging systems in which other bottlenecks (e.g., computation or main memory bottlenecks) get alleviated. 
For example, while metagenomics can benefit from near-data processing at the main memory level, i.e., processing-in-memory (PIM)~\cite{wu2021sieve,shahroodi2022krakenonmem,shahroodi2022demeter,hanhan2022edam,zou2022biohd}, these approaches still incur the overhead of moving the large, low re-use data from storage. In fact, by accelerating other overheads, the impact of storage I/O on end-to-end performance increases. For example, for the 0.3-TB and 0.6TB Kraken2 databases, 
and considering a state-of-the-art PIM accelerator~\cite{wu2021sieve} of Kraken2,  \dram's end-to-end performance is on average 26.1$\times$ (3.0$\times$) better than \ssdc (\ssdp), respectively. We conclude that while PIM can provide significant benefits for metagenomics analysis, it does not alleviate the overhead of moving large, low-reuse data from the storage device.

\subsection{Our Goal}

Processing data directly in the storage system, where the data originally resides, via ISP can be a fundamental and cost-efficient approach for reducing the overheads of moving large volumes of low-reuse data across the system.  ISP provides three key benefits. First, it reduces unnecessary data movement from the storage system. Second, it reduces the overall performance and energy burden of the application from the rest of the system, freeing them up to do other useful work meanwhile. Third, as shown by many prior works (e.g.,~\cite{mailthody2019deepstore,kang2021mithrilog,koo2017summarizer,mansouri2022genstore}), ISP can benefit from the SSD's larger internal bandwidth. 
For example, with 8 (16) channels for \ssdc (\ssdp) and the maximum per-channel bandwidth of 1.2~GB/s, the maximum internal bandwidth is calculated to be 9.6~GB/s (19.2~GB/s).
The external sequential-read bandwidth of \ssdc (\ssdp) is 560~MB/s  (7~GB/s). 
Therefore, the internal bandwidth is 17.4$\times$ (2.7$\times$) larger than the external in \ssdc (\ssdp).

Despite its benefits, designing an ISP system for metagenomics is challenging because none of the existing approaches can be directly implemented as an ISP system effectively due to the constrained hardware resources in the SSD. Techniques such as \randomio hinder leveraging ISP’s large potential by preventing the full utilization of the SSD’s internal bandwidth due to costly conflicts in internal SSD resources~\cite{nadig2023venice,tavakkol2018flin,kim2022networked} caused by random accesses. Techniques such as \streamio predominantly incur more suitable streaming accesses, but they do so at the cost of more computation and main memory capacity requirements, posing challenges for ISP.
Therefore, directly adopting these approaches in the storage system incurs performance, energy, and storage device lifetime overheads. 
\textbf{Our goal} is to improve the performance of the end-to-end metagenomic analysis workflow by reducing data movement overhead, right from the storage system, in a cost-effective manner and by addressing the challenges of ISP.

\section{\proposal}
\label{sec:mechanism}

We propose \proposal, the first ISP system for metagenomics.
\proposal extends the existing SSD controller and FTL 
and integrates into the system 
so that when not used for ISP, the SSD can be accessible for all other applications as it would be in a conventional system.
We address the challenges of ISP for metagenomics via hardware/software co-design to enable \emph{cooperative ISP} via an efficient pipeline between the host \js{and the storage system} to \js{maximally leverage and orchestrate their capabilities}. 
Based on our detailed analysis of metagenomic tasks, we partition the tasks and map each part to the most suitable system, while ensuring that the partitioning does not lead to additional overhead due to data transfer time. 
\proposal efficiently coordinates the data/control flow between the host and storage systems.
For each part, we develop algorithmic optimizations, data mapping schemes, and/or lightweight in-storage accelerators based on the characteristics of the task, real datasets, and the underlying hardware.

We explain \proposal's \emph{Step 1: Preparing the Input Queries}, \emph{Step2: Finding Candidate Species}, and the optional \emph{Step 3: Abundance Estimation}, and \proposal-FTL in \sects{\ref{sec:mech-stage1}, \ref{sec:mech-stage2}, \ref{sec:mech-stage3}, and \ref{sec:mech-ftl}}, respectively. 
\sect{\ref{sec:mech-overview}} shows the end-to-end walkthrough. 
\sect{\ref{sec:mech-multi-sample}} shows optimizations for use cases where multiple samples require the same database simultaneously.

\subsection{Step 1: Preparing the Input Queries}
\label{sec:mech-stage1}

\proposal works with lexicographically-sorted data structures to avoid expensive random accesses to the SSD (similar to \streamio described in \sect{\ref{sec:background-metagenomics}}). Similar to many other metagenomic tools (including ~\cite{wood2014kraken,shahroodi2022krakenonmem,lapierre2020metalign,truong2015metaphlan2}), we assume the sorted k-mer databases are pre-built before the analysis. However, sorting k-mers extracted from the input query read set (i.e., sample) cannot efficiently be done offline because it requires storing a separate large data structure (i.e., sorted k-mer set) alongside each sample, which could potentially be even larger than the original sample and lead to a significant waste of storage capacity. Therefore, to prepare the input queries, \proposal 1) extracts k-mers from the sample, 2) sorts the k-mers, and if needed, 3) prunes some k-mers with user-defined criteria.

In \proposal's pipeline, we execute this step in the host for three reasons. First, sorting benefits from the relatively larger DRAM and compute resources in the host. Second, due to the host's large DRAM, performing this step in \gram{the} host leads to significantly fewer writes to the flash chips and can positively impact their lifetime. For typical metagenomic read sets, storing k-mers extracted from reads within a sample takes tens of gigabytes (e.g., on average 60 GB in our experiments with standard CAMI read sets~\cite{meyer2021critical}, as detailed in \sect{\ref{sec:methodology}}). While it is possible to generate and sort these k-mers inside the SSD by leveraging the small internal DRAM (e.g., 4-GB DRAM for a 4-TB SSD), these operations would require frequent writes to the flash chips. Third, by effectively leveraging the host for this step, we can enable pipelining and overlapping the execution of Steps 1 and Step 2 (which requires searching the terabyte-scale low-reuse metagenomic database).

To enable the efficient execution of this step in the host, we need to ensure two points. First, partitioning the application between the host system and the SSD in different stages should not incur significant overhead due to data transfer time. 
Second, while it is reasonable in most cases to expect the host DRAM to be large enough to contain all extracted k-mers from a sample, \proposal should support scenarios where this is not the case and minimize the performance, lifetime, and endurance overheads of writes due to page swapping.

\subsubsection{K-mer Extraction}
\label{sec:mech-stage1-kmer-extraction}

\begin{figure}[t]
    \centering
    \includegraphics[width=.97\linewidth]{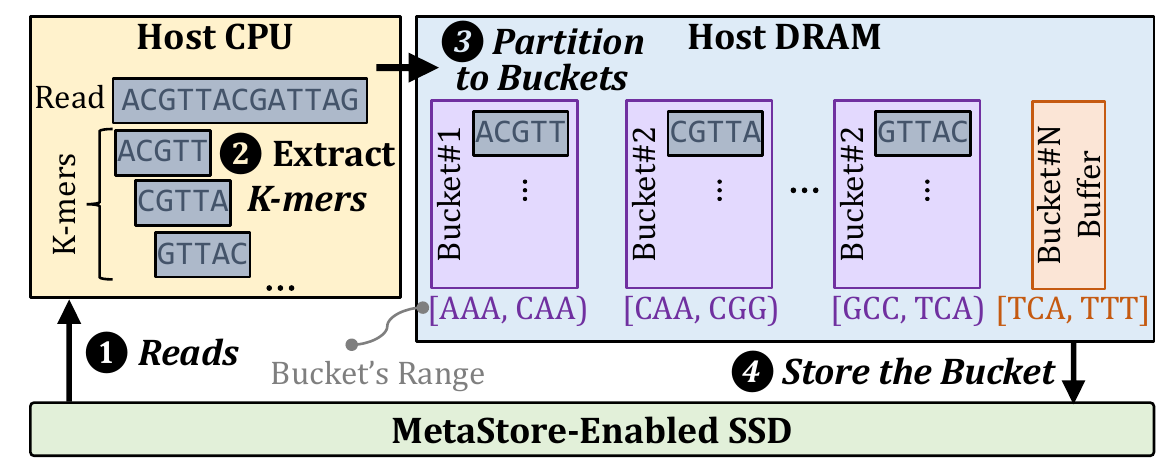}
    \caption{K-mer extraction.}
    \label{fig:kmer-gen}
\end{figure}

To reduce data transfer overhead between partitions, we propose a new input processing scheme by improving upon input processing in KMC~\cite{kokot2017kmc3}. We partition the k-mers into buckets, where each bucket corresponds to a lexicographical range and stores k-mers falling into that range. This way, we can overlap the k-mer sorting and transfer of one bucket with ISP on the previous buckets. Step 2 (\sect{\ref{sec:mech-stage2}}) can start working on a bucket as soon as that bucket's k-mers are sorted and pruned. This is because the database k-mers are also sorted and can already be accessed within the range corresponding to the bucket. \fig{\ref{fig:kmer-gen}} shows an overview of \proposal's k-mer extraction. The host reads the input reads (\circled{1} in \fig{\ref{fig:kmer-gen}}), extracts their k-mers (\circled{2}), and stores them in the buckets (\circled{3}).\footnote{To avoid bucket size imbalance, we generate preliminary buckets for a small subset of k-mers at the beginning. If we observe an imbalance between these buckets, we balance the buckets by merging some to meet a user-defined bucket count (default 512).}

To support scenarios where the extracted k-mers of one sample do not fully fit in the host DRAM, \proposal pins some buckets to DRAM (e.g., Buckets $1$ to $N-1$ in \fig{\ref{fig:kmer-gen}}) and uses the SSD to store the others. This way, the k-mers belonging to the DRAM-pinned buckets do not move back and forth between the host and the SSD (\circled{4}). To further reduce the overhead of writes/reads for the SSD-pinned buckets, \proposal's bucketing scheme takes two measures. First, it uses buffers in the host DRAM for the SSD-pinned buckets so that when full, it \js{flushes} the contents of the buffers to the SSD, utilizing the \js{full sequential-write} bandwidth. Second, we map the bucket's k-mers across SSD channels, dies, and planes evenly for parallelism and in an interleaved manner. Since \proposal does not require writes to the flash chips \new{after this step}, it can flush all \gram{of} the FTL metadata for write-related management to free up internal DRAM for the next steps \new{(details in \sect{\ref{sec:mech-ftl}})}. 

\subsubsection{Sorting}
\label{sec:mech-stage1-sorting}
\hm{After generating all k-mer buckets,}
\proposal proceeds to sort the k-mers within the individual buckets. As soon as the sorting of a specific bucket (i.e., bucket $i$) is completed, \proposal transfers this bucket to the DRAM inside the SSD in batches (to go through the second stage, as described in \sect{\ref{sec:mech-stage2}}). 
Meanwhile, during the transfer of bucket $i$, \proposal advances to the sorting of bucket $i+1$. 
 \proposal can also orthogonally use a sorting accelerator to perform sorting.

\subsubsection{Excluding K-mers}
\label{sec:kmer-exclusion}
\proposal, like various tools~\cite{benoit2016multiple,bovee2018finch}, can exclude k-mers based on user-defined frequencies. Users can exclude 1)~overly common k-mers, which are not discriminative, and 2)~k-mers that appear only once, which may represent sequencing errors or low-abundance organisms that are hard to distinguish from random occurrences.\footnote{Other k-mer selection/sampling criteria can also be orthogonally and flexibly integrated in this step.} Excluding these k-mers can improve accuracy and reliability. Exclusion occurs after sorting, where k-mers are inherently counted, and before transferring buckets to Step 2. While the size of the extracted query k-mers (\sect{\ref{sec:mech-stage1-kmer-extraction}}) can be large (on average 60~GB in our experiments), the size of the k-mer set selected to go to the next stage is much smaller (on average 6.5~GB) and is also significantly smaller than the database that can be up to several terabytes~\cite{rautiainen2023telomere,jarvis2022semi}.

\subsection{Step 2: Finding Candidate \rev{Species}}
\label{sec:mech-stage2}

In this stage, \proposal finds the \hm{species} present in the sample
by 1) intersecting the query k-mers and the database,
 and 2) finding the tax IDs of the intersecting k-mers.
We perform this stage inside the SSD since it requires streaming the large database with low reuse and involves only lightweight computation. 
This enables \proposal to leverage the large SSD-internal bandwidth 
 and alleviate the overall burden of moving/analyzing large, low-reuse data from the rest of the system.

Considering the SSD hardware limitations, \proposal should leverage the full internal bandwidth while ensuring it does not require expensive hardware resources inside the SSD (e.g., large DRAM bandwidth and costly logic units). Performing this step effectively inside the SSD requires efficient coordination between the SSD and the host, mapping, hardware design, and storage technology-aware algorithmic optimizations.

\subsubsection{Intersection Finding}
\label{sec:mech-stage2-1}

In this step, \proposal finds the intersecting k-mers, i.e., k-mers present in both the query k-mer buckets arriving from the host 
and the large k-mer database stored in the flash chips.

Relying on the internal DRAM for 1) buffering the query k-mers arriving from the host and the database k-mers arriving from the SSD channels at full bandwidth and 2) streaming through both to find their intersection can pressure the bandwidth of the internal DRAM. For example, reading at the full internal bandwidth in a high-end SSD can already exceed the LPDDR4 DRAM bandwidth used in current SSDs~\cite{zou2022assasin, samsung980pro,samsungPM1735} and even the \mbox{16-GB/s} DDR4 bandwidth~\cite{zou2022assasin,ddr4sheet}. To address this challenge, we adopt an approach similar to \cite{zou2022assasin} and perform computation directly on flash data streams without needing to write them to the internal DRAM. Despite its key benefits, this approach requires large buffers (64KB for input and 64 KB for output, \emph{per SSD channel}). 

To enable computation directly on the flash data streams at a low cost, we leverage two key features of \proposal to find the minimum required buffer size. First, the computation in this step is very lightweight and does not require a large buffer for data awaiting computation. Second, data is uniformly distributed across channels, and each comparator needs to only operate on the data received from one channel. 
Based on these, we directly read data from the flash controller and include \emph{two} k-mer registers per channel. While one register stores a k-mer as the input to the comparator, the other stores the bytes in the next k-mer as they arrive from the flash controller. 
For the cases where \proposal cannot read the next database k-mer (i.e., when the comparator needs to compare database k-mer $i$ with several k-mers from the query bucket), \proposal will not fetch the next k-mer from the flash controller. 
This way, by only using two registers, \proposal can directly compute on the flash data stream at a low cost, and without bottlenecking the limited internal DRAM bandwidth. 

\begin{figure}[t]
    \centering
    \includegraphics[width=\linewidth]{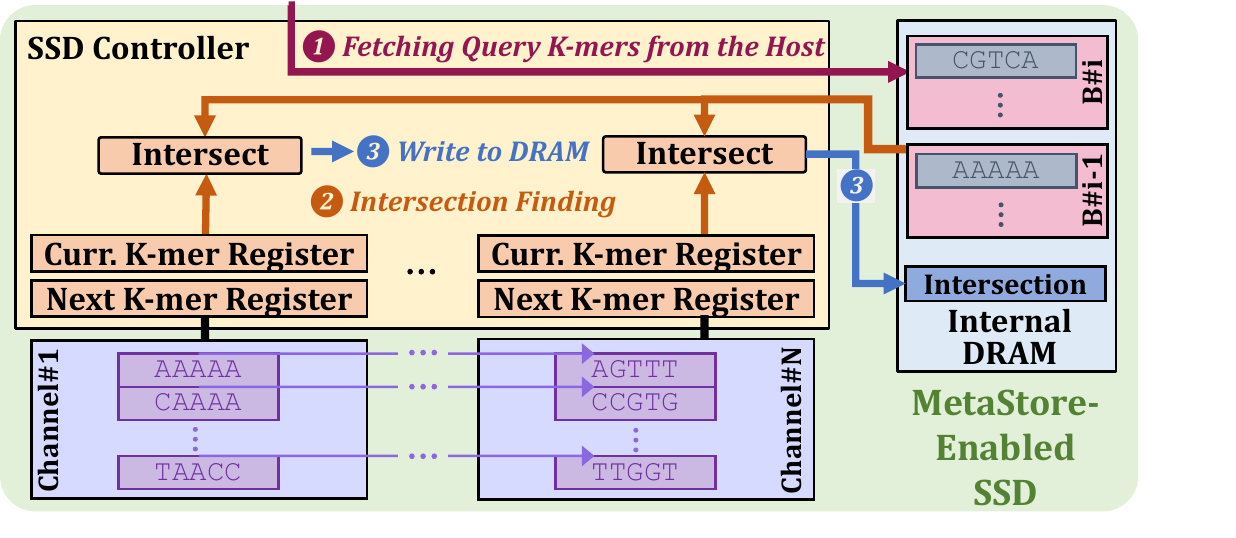}
    \caption{Intersection finding.}
    \label{fig:intersection-finding}
\end{figure}

\fig{\ref{fig:intersection-finding}} shows the operations.
First, \proposal reads the query k-mers from the host in batches to the internal DRAM (\circled{1} in \fig{\ref{fig:intersection-finding}}). Second, it concurrently reads both the sorted query k-mers from the internal DRAM and the sorted database k-mers from all flash chips. It performs a comparison operation to find their intersection (\circled{2}). To do so, \proposal uses per-channel comparators located on the SSD controller to operate on each channel's data stream. Third, \proposal writes the intersecting k-mers to the internal DRAM for further analysis (\circled{3}).

\head{Fetching Query K-mers}
To leverage the external bandwidth efficiently, \proposal moves buckets from \gram{the} host to \gram{the} internal DRAM in batches, with a size equal to the size of the data that can be sent from the host to the SSD. We consider two batches of k-mers stored in the internal DRAM.
While transferring batch $B\#i$, \proposal performs intersection finding on $B\#i-1$. 
For an SSD with 8 channels, 4 dies/channel, 2 planes/die, and 16-KiB pages, \proposal only requires space for two 1-MiB batches in the internal DRAM.

\head{Intersection Finding}
\proposal reads the query k-mers from the internal DRAM and the database k-mers from the flash controller. This step runs in a pipeline with the previous. We store the database evenly across different channels to leverage the full internal bandwidth when sequentially reading data using multi-plane operations.  \proposal finds the intersecting k-mers using a simple comparator logic. 
If a database k-mer equals a query k-mer, we record it as an intersection in the internal DRAM. If a query k-mer is larger (smaller), \proposal reads the next database (query) k-mer. MetaStore’s Control Unit, located on the SSD controller, receives the comparison results and issues the control signals accordingly.\footnote{The figures exclude the Control Unit and its connections for readability.}

\head{Storing the Intersecting K-mers}
\proposal stores the intersecting k-mers in the internal DRAM.\footnote{The intersection does not have a strict size requirement and can fill in the remaining free space of the internal DRAM opportunistically. Typically, due to its small size, the intersection k-mer set can fit within the internal DRAM. However, in a corner case where the intersecting k-mers cannot fully fit in the internal DRAM, \proposal can start the tax ID retrieval (\sect{\ref{sec:mech-stage2-2}}) for the already-found intersecting k-mers. After that, it can return to this step and over-write the old intersecting k-mers in DRAM.} The internal DRAM's bandwidth needs to support  1) fetching the query k-mers from the host, 2) reading them out for comparison, 3) storing the intersection, and 4) FTL metadata. Since the query k-mer set and FTL metadata (FTL details in \sect{\ref{sec:mech-ftl}}) are significantly smaller than the database, they can be accessed at a much smaller bandwidth than reading the database. 
For example, for our datasets (\sect{\ref{sec:methodology}}), while fully leveraging the internal bandwidth of \ssdp, \proposal requires only 2.4~GB/s of DRAM bandwidth. 

\rev{Leveraging our optimizations, \proposal's ISP requires only simple computation and small buffers. This enables \proposal's ISP to also run on the embedded cores on the SSD controller. We evaluate both \proposal configurations (with the lightweight accelerator or embedded cores) in \sect{\ref{sec:eval}}}. Choosing between these two \proposal configurations is a design decision with different trade-offs, with the accelerator configuration achieving better performance and significantly higher power efficiency.

\subsubsection{Retrieving Tax IDs}
\label{sec:mech-stage2-2}

This step retrieves the species tax IDs of the intersecting k-mers. To do so, we compare the intersecting k-mers against the pre-built k-mer sketches representing each species' reference genomes. Similar to \cite{lapierre2020metalign}, we use CMash~\cite{liu2022cmash} to generate sketches. \proposal can orthogonally use other sketches. \proposal flexibly supports variable-sized k-mers.  As shown by prior works~\cite{liu2022cmash,kim2016centrifuge}, while longer k-mers offer greater discrimination, they may result in incomplete detection of hits between queries and sketches. In such cases, users may also try smaller k-mers (i.e., k-mers prefixes) to identify additional matches.

\begin{figure}[b]
    \centering
    \includegraphics[width=0.9\linewidth]{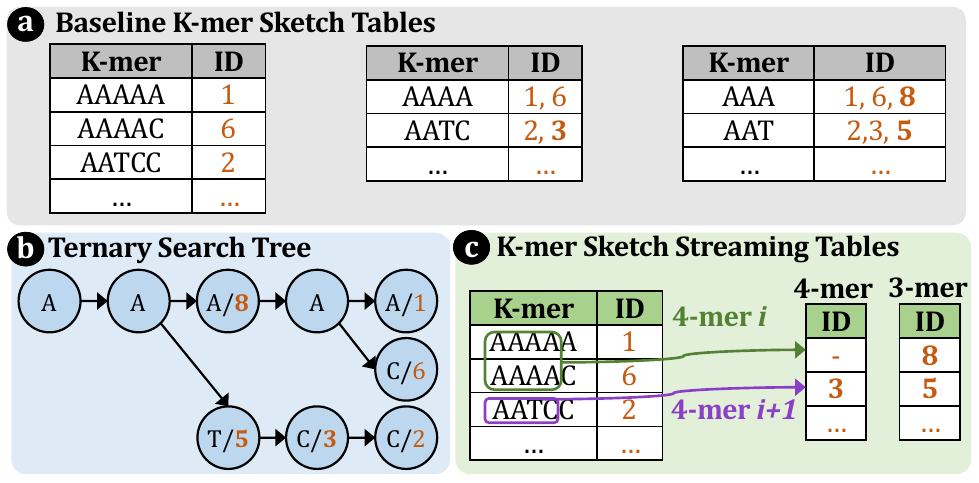}
    \caption{Sketch data structures.}
    \label{fig:taxid-structures}
\end{figure}

Finding tax IDs for variable-sized k-mers is challenging since it requires many pointer-chasing operations on a large data structure that may not fit in the internal DRAM.
To find the tax IDs of k-mers, some approaches~\cite{marchet2021data,liu2022cmash,lapierre2020metalign} devise data structures to encode the k-mer information in a space-efficient manner.
For example, CMash~\cite{liu2022cmash} encodes k-mers \gram{of} variable 
sizes in a ternary search tree. \fig{\ref{fig:taxid-structures}} shows 5-, 4-, and 3-mers along\rev{side} 
their tax IDs in \circled{a} separate tables, as used by some prior approaches~\cite{koslicki2016metapalette,weging2021taxonomic}, and \circled{b} in a ternary search tree. 
This tree structure is devised to 1) save space and 2) retrieve the tax IDs for all k-mers with $k \leq k_{max}$ that are prefixes of \hm{a query} $k_{max}$-mer. For example, as shown in \fig{\ref{fig:taxid-structures}}, when traversing the tree to look up the 5-mer \texttt{AATCC}, we can look up the 4-mer \texttt{AATC} during the same traversal. 
Despite its benefits, this approach requires up to $k_{max}$ pointer-chasing \gram{operations} for \emph{each lookup}. Performing these operations inside the SSD is challenging since the tree can be larger than the internal DRAM, and pointer chasing on flash arrays is expensive due to their significantly larger latency compared to DRAM.

While \proposal can execute this part in the host, we identify a new optimization opportunity leveraging unique features of ISP (i.e., large internal bandwidth and storage capacity),
which avoids pointer-chasing at the cost of larger data structures. 
\fig{\ref{fig:taxid-structures}} \circled{c} shows an overview of our approach, \emph{K-mer Sketch Streaming (KSS)}. 
For k-mers with size $k_{max}$, \proposal stores the k-mer sketches and their tax IDs similar to \circled{a}. \proposal sorts this table alphabetically. 
For smaller values of $k$, \proposal uses the prefixes of the \mbox{$k_{max}$-mers} to retrieve smaller k-mers. For each smaller k-mer, it only stores the tax IDs that are \emph{not} attributed to their corresponding larger k-mer.\footnote{A smaller k-mer can be found in more \hm{species} since it is less unique than its corresponding $k_{max}$-mers.} Storing the k-mer itself is not needed because it can be retrieved as the prefix of the larger k-mers.
This encoding allows for tax ID retrieval by sequentially streaming through the intersecting k-mers  (which are already sorted) and these \cmashopt tables. 
While \gram{the} \cmashopt data structure is larger than \circled{b}, it is suitable for ISP due to the high internal bandwidth and large capacity. \cmashopt can also be helpful for outside-SSD execution with SSDs with \hm{high} external bandwidth (\sect{\ref{sec:eval-main}}). \cmashopt leads to 7.5$\times$ smaller data structures compared to the 107-GB \circled{a}, and 2.1$\times$ larger compared to \circled{b} (dataset details in \sect{\ref{sec:methodology}}).

\begin{figure}[h]
    \centering
    \includegraphics[width=.97\linewidth]{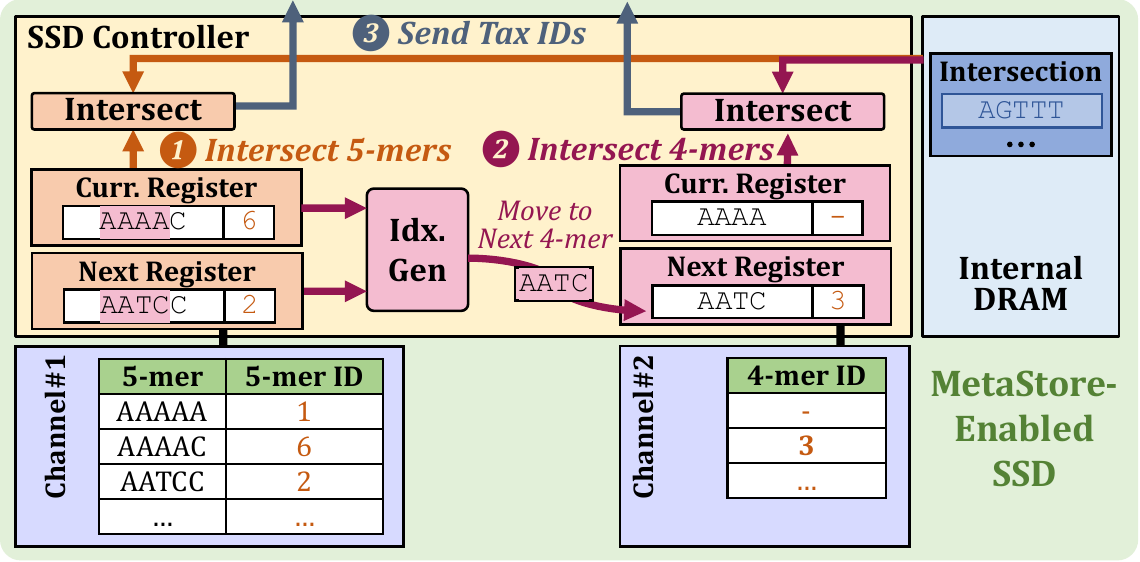}
    \caption{Tax ID retrieval.}
    \label{fig:taxid-finding}
\end{figure}

\fig{\ref{fig:taxid-finding}} shows \gram{an} overview of \proposal's Tax ID retrieval. As an example, we demonstrate retrieving 5-mers and 4-mers. First, \proposal reads the intersecting k-mers (i.e., 5-mers) from the internal DRAM and concurrently reads the 5-mer sketches and their IDs from an SSD channel. It performs a comparison operation on the 5-mer streams to find matches (\circled{1}). 
Second, to find 4-mer matches, \proposal compares the \emph{prefixes} of the intersecting 4-mers with the prefixes of the 4-mer sketches (\circled{2}). 
\proposal devises a lightweight \emph{Index Generator}. I\rev{t} compares the 4-mer prefixes of each pair of consecutive 5-mer sketches, and when the prefixes are not equal (i.e., a new 4-mer appears), it points to the next element in the 4-mer table.  We map the data related to 5-mers and 4-mers in different channels to allow for concurrent accesses to leverage the full internal bandwidth.
Third, \proposal sends the retrieved tax IDs to the host (\circled{3}) as the IDs of the candidate species present in the sample.

\subsection{Step 3: Abundance Estimation}
\label{sec:mech-stage3}

For metagenomic applications that require abundance estimation, \proposal integrates further analysis on the candidate species identified as present at the end of Step 2. \proposal can flexibly integrate with different approaches used in various tools,
such as \rev{\inum{i}}~lightweight statistics~\cite{lu2017bracken,dimopoulos2022haystac}
or \inum{ii}~more accurate and costly read mapping~\cite{lapierre2020metalign,kim2016centrifuge,milanese2019microbial}, where the input read set is mapped to the reference genomes of candidate present species. In this step, \proposal facilitates integration with different approaches. The lightweight statistical approaches work directly on the output of Step 2. To facilitate the more complex read mapping, \proposal prepares the data needed by the software or hardware mapper used in the system.  The mapper needs the query reads and an index of the reference genomes of the candidate species, commonly used during read mapping~\cite{li2018minimap2}. 
Constructing individual species indexes can be a one-time offline task. However, generating a unified index for the collection of the present species cannot be done offline, as these specific species are undetermined beforehand.

\begin{figure}[h]
    \centering
    \includegraphics[width=0.8\linewidth]{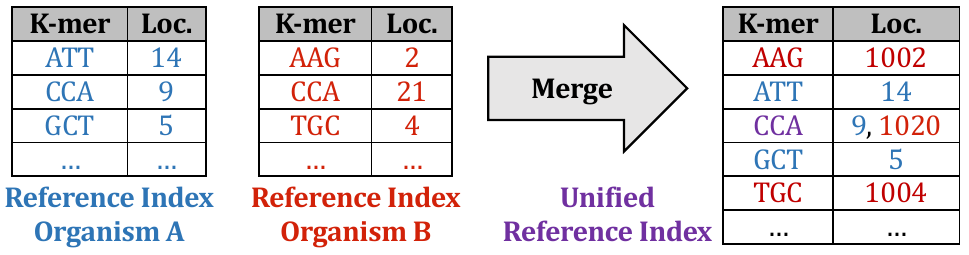}
    \caption{Merging the reference indexes.}
    \label{fig:index-merger}
\end{figure}

We facilitate the read mapping process by generating a unified reference index of the species detected in Step 2, inside the SSD directly while reading the individual indexes of these species. \fig{\ref{fig:index-merger}} demonstrates this with an example of two pre-built and pre-sorted indexes of species A and B. Each index entry shows a k-mer and its location in that species' reference genome. \proposal reads each index sequentially and merges them. When finding a common k-mer between the two indexes, (e.g., \texttt{CCA} in \fig{\ref{fig:index-merger}}),  it stores both locations for that k-mer in the unified index. In this example, assuming the size of the reference genome  A is 1000, we store the location of \texttt{CCA} in B as 1020 in the unified index.

\subsection{\proposal-FTL}
\label{sec:mech-ftl}

\proposal-FTL needs simple changes to the baseline FTL, to handle communication/data flow between the host and the SSD.
At the beginning, \proposal-FTL maintains all metadata of the regular FTL in the internal DRAM.
For the only step that requires writes (\sect{\ref{sec:mech-stage1-kmer-extraction}}, at the beginning of Step 1 in the host), \proposal-FTL still uses the write-related metadata (e.g., L2P, bad-block information). 
After this step, \proposal does not require writes, so it flushes the regular L2P and loads \proposal's L2P, while still keeping the other metadata of a regular FTL.

\begin{figure}[t]
    \centering
    \includegraphics[width=\linewidth]{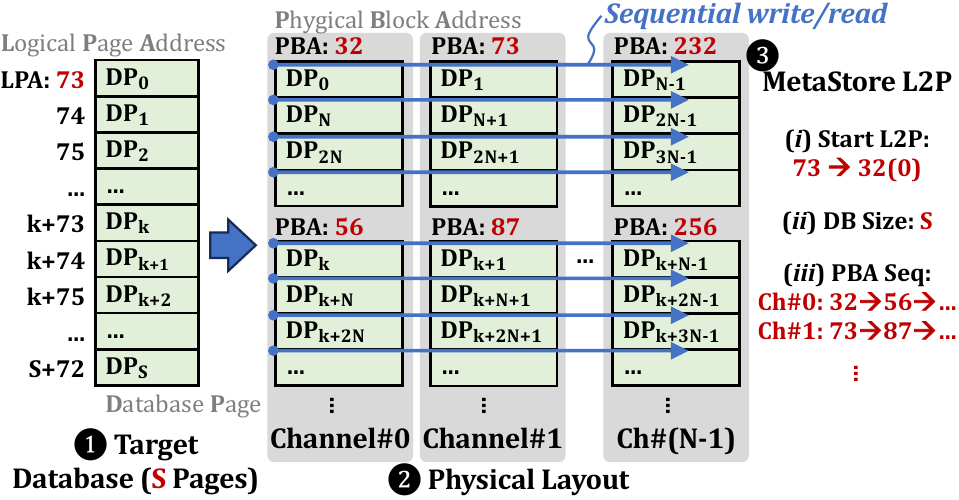}
    \caption{Data layout and mapping data structure in \proposal.}
    \label{fig:ms-ftl}
\end{figure}

\fig{\ref{fig:ms-ftl}} shows how \proposal-FTL manages the target data stored in NAND flash memory, i.e., databases\footnote{K-mer and sketch databases are the only data structures accessed from NAND flash memory during \proposal's ISP operations. The read set k-mers get transferred from the host system to the SSD during ISP.}, with a reduced amount of metadata.
When initially storing a database (\circled{1} in \fig{\ref{fig:ms-ftl}}), \proposal-FTL \circled{2} \emph{evenly and sequentially} distributes the data across all SSD channels while ensuring that every \emph{active} block~\cite{kawaguchi1995flash} (which is currently used for writing) in different channels has the same page offset.\footnote{This can be done by either allocating dedicated blocks for the database.}
Doing so allows \proposal-FTL to only keep a small mapping data structure (\circled{3}) that consists of \inum{i} the mapping between start logical page address (LPA) and physical page address (PPA), \inum{ii} the database size, and \inum{iii} the sequence of physical block addresses (PBAs) used to store the database, instead of the entire LPA-to-PPA mappings.
This is because \proposal always accesses the database \emph{sequentially}. 
As shown in \fig{\ref{fig:ms-ftl}}, \proposal-FTL can sequentially read the stored database from the starting LPA while performing page reads in a round-robin manner across channels; it just increments the PPA within a physical block and resets the PPA when reading the next block.
Compared to the regular L2P, whose space overhead is 0.1\% of stored data (4 bytes per 4 KiB), \proposal's L2P is very small. 
For example, \proposal only requires $\sim$1.3 MB to store a 4-TB database, assuming a physical block size of 12~MB: 4~bytes for each of the 349,525 (4 TB/12 MB) used blocks (and a few bytes for the start L2P mapping and database size).
Note that the only metadata other than L2P mappings that must be kept during ISP is the per-block access count for read-disturbance management~\cite{cai2015read}, so the size of total metadata for \proposal is up to 2.6~MB for a 4-TB SSD. Flushing regular L2P into flash memory and using this smaller L2P enables us to exploit most of the internal DRAM during ISP (to store query batches $B\#i$ and $B\#i-1$ and Intersection, as shown in \figs{\ref{fig:intersection-finding} and \ref{fig:taxid-finding}}).

\rev{\proposal's in-storage accelerators are located on the SSD controller and access data \emph{after} ECC (which is designed to match the channels bandwidth).}
\proposal performs other tasks necessary for ensuring reliability (e.g., refreshing data to prevent uncorrectable errors\cite{cai-hpca-2017, luo2018improving,luo-hpca-2018,cai2015read,cai2013error, cai2012flash, cai2017error, ha2015integrated, cai-insidessd-2018,luo2015warm}) \emph{before or after} the ISP phase. This is because 1) the duration of each \proposal process is significantly smaller than the manufacturer-specified threshold for reliable retention age (e.g., one year~\cite{micron3dnandflyer}), and 2) since \proposal only reads data sequentially with low re-use, it can avoid read disturbance errors~\cite{cai2015read} during ISP.

\subsection{End-to-End Walkthrough}
\label{sec:mech-overview}

\begin{figure}[t]
    \centering
    \includegraphics[width=\linewidth]{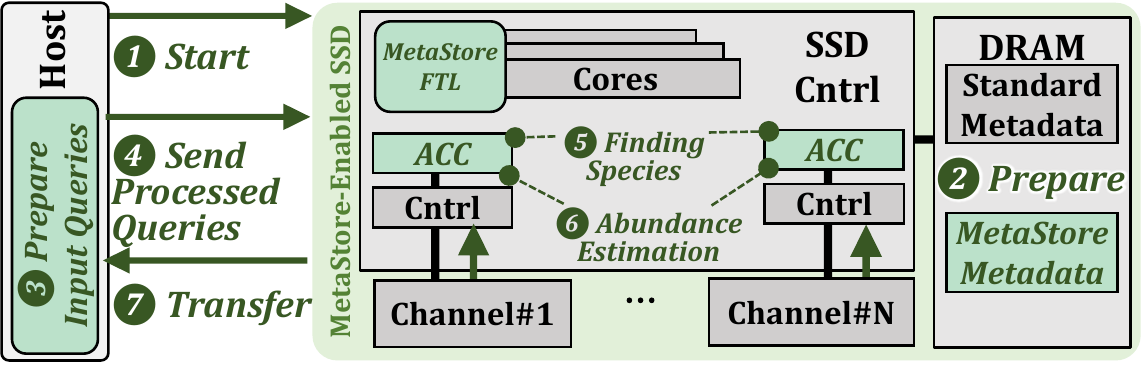}
    \caption{Overview of \proposal.}
    \label{fig:metastore_overview}
\end{figure}

\fig{\ref{fig:metastore_overview}} shows \hm{an} end-to-end walkthrough of \proposal's steps.
Upon receiving \gram{a} notification from the host to initiate metagenomic analysis \circled{1}, \proposal readies itself for collaborative ISP (\circled{2}).
To do so, \proposal loads the \proposal-FTL metadata needed for its operations.
After the preparation, \proposal starts the  \hm{three-step} execution. 
In \rev{\textbf{Step 1}, the host processes the input read queries (\circled{3})} and transfers them in batches to the SSD (\circled{4}). 
In \textbf{Step 2}, the ISP units find the species present in the sample (\circled{5}). Steps 1 and 2 run in a pipelined manner for consecutive query batches.
In \textbf{Step 3}, \proposal facilitates the integration with different abundance estimation approaches by preparing (\circled{6}) and transferring (\circled{7}) the data needed for any further analysis.
With its hardware/software co-design, \proposal avoids channel conflicts and frequent management tasks (by not requiring writes) during ISP, and leverages the full internal bandwidth.

\new{\subsection{Multi-Sample Analysis}
\label{sec:mech-multi-sample}

\newcommand\bases{\textsf{Base-S}\xspace}
\newcommand\basem{\textsf{Base-M}\xspace}
\newcommand\mss{\textsf{MS-S}\xspace}
\newcommand\msm{\textsf{MS-M}\xspace}
\newcommand\optm{\textsf{Opt-M}\xspace}

For some use cases, 
a metagenomic study can have multiple read sets (i.e., samples) available at the same time that need to access the same database.
If the host's DRAM is larger than the k-mer sizes extracted from a sample, we use the available DRAM opportunistically to buffer k-mers extracted from \emph{several} samples. This way, \proposal streams through the database only once.
\fig{\ref{fig:multi-timeline}} shows the timeline of analyzing \gram{a} single (\textsf{S}) or multiple (\textsf{M}) samples in the baseline (\textsf{Base}), in our proposed optimized approach in software (\textsf{Opt}), and in \proposal (\textsf{MS}).
\proposal can flexibly integrate with a sorting accelerator to further improve end-to-end performance.

\begin{figure}[h]
    \centering
    \includegraphics[width=\linewidth]{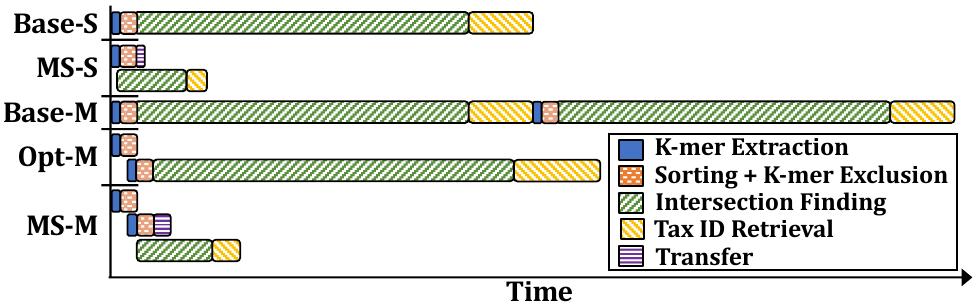}
    \caption{\new{Timeline of multi-sample analysis.}}
    \label{fig:multi-timeline}
\end{figure}
}

\section{Evaluation Methodology}
\label{sec:methodology}

\head{SSDs}  Our system is equipped with \ssdc~\cite{samsung870evo} and \ssdp~\cite{samsungPM1735} as described in \sect{\ref{sec:motivation-ovhd}}.

\head{Performance} 
We design a simulator that models all \proposal's components, including host operations, accessing flash chips, the internal DRAM, the in-storage accelerator, and interfaces between the host and the SSD. We feed the latency and throughput of each of \proposal components to this simulator.  
For the components in the \textbf{hardware-based} steps (e.g., ISP units in Steps 2 and 3):  We implement \proposal's logic components in Verilog. We synthesize them using the Synopsys Design Compiler~\cite{synopsysdc} with a 65~nm library\cite{umcL65nm} to estimate their frequency. We use two state-of-the-art 
simulators, Ramulator~\cite{kim2016ramulator, ramulatorsource} for DRAM to model SSD's internal DRAM accurately, and MQSim~\cite{tavakkol2018mqsim} for SSD's internal operations.
For the components in the \textbf{software-based} steps (e.g., host operations in Step 1): We measure performance on a real system, an AMD$^\text{\textregistered}$ EPYC$^\text{\textregistered}$ 7742 CPU with 128 physical cores, 1-TB DRAM (in all experiments unless stated otherwise), and the mentioned SSDs.  We will open-source our simulator and scripts. 
For the software baselines, we measure performance on the mentioned real system, using the number of threads that lead to each tool's best performance.

\head{Area and Power} 
For logic components, we use the results from our Design Compiler synthesis. 
We implement \proposal's logic components in Verilog. 
For SSD power, we use the values of a Samsung 3D NAND flash-based SSD~\cite{samsung860pro}. For DRAM power, we base the values on a DDR4 model~\cite{ddr4sheet, ghose2019demystifying}. For the CPU cores, we use AMD$^\text{\textregistered}$ \textmu{}Prof~\cite{microprof}.

\head{Baseline Metagenomic Tools} We use a state-of-the-art performance-optimized (\textsf{P-Opt}) tool, Kraken2 + Bracken~\cite{wood2019improved}, and a state-of-the-art accuracy-optimized (\textsf{A-Opt}) tool, Metalign~\cite{lapierre2020metalign}. 
Particularly, for the presence/absence task, we use Kraken2 without Bracken, and Metalign without mapping (i.e., only KMC~\cite{kokot2017kmc3} + CMash~\cite{liu2022cmash}). 
For abundance estimation, we use Kraken2 + Bracken, and full Metalign.
For both Metalign and \proposal, we use GenCache~\cite{nag2019gencache} for mapping. We use the optimal mapping throughput as reported by the original paper~\cite{nag2019gencache}.
 \proposal can flexibly integrate with other mappers. 
We also compare \proposal against a state-of-the-art PIM k-mer matching accelerator, Sieve~\cite{wu2021sieve}, \hm{for} accelerat\hm{ing} Kraken2's pipeline. We use the optimal k-mer matching performance as reported by the original paper~\cite{wu2021sieve}.

\head{Datasets} 
We use three query read sets from the commonly-used CAMI benchmark~\cite{sczyrba2017critical}, with low/medium/high \hm{genetic} diversity. Each read set has 100 million reads.
We generate a database based on \hm{microbial genomes drawn from NCBI's databases}~\cite{ncbi2020,lapierre2020metalign} using default parameters for each tool. For Kraken2~\cite{wood2019improved}, this results in a 293~GB database. For Metalign~\cite{lapierre2020metalign}, this results in a 701~GB k-mer database
and \gram{a} 6.9~GB sketch tree.
\proposal uses the same 701~GB k-mer database and a 14~GB sketch database for our \cmashopt.

\section{Evaluation}
\label{sec:eval}

\subsection{Presence/Absence Identification Analysis}
\label{sec:eval-main}

\newcommand\popt{\textsf{P-Opt}\xspace}
\newcommand\aopt{\textsf{A-Opt}\xspace}
\newcommand\msnol{\textsf{MS-NOL}\xspace}
\newcommand\msext{\textsf{Ext-MS}\xspace}
\newcommand\mscc{\textsf{MS-CC}\xspace}
\newcommand\msfull{\textsf{MS}\xspace}

\begin{figure}[h]
\centering
 \includegraphics[width=\linewidth]{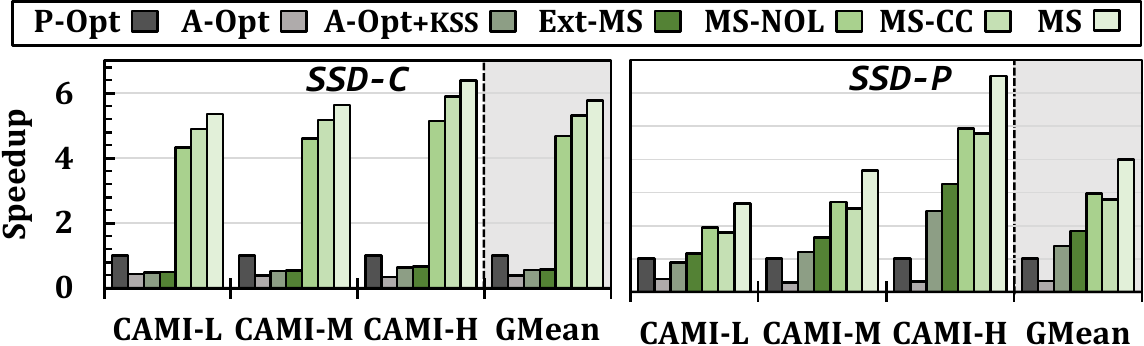}
\caption{\rev{Speedup for different SSDs and input sets.}}
\label{fig:main-eval}
\end{figure}

We use 1\gram{-}TB DRAM in this analysis (smaller than all datasets used here). 
\fig{\ref{fig:main-eval}} shows the performance of five tools: 
1)~\popt,
2)~\aopt,
3)~{\aopt}+\cmashopt: \aopt that leverages the software implementation of our \cmashopt approach (\sect{\ref{sec:mech-stage2-2}}), instead of Metalign's CMash~\cite{lapierre2020metalign}, 
4)~\msnol: a \proposal implementation without overlapping the host and SSD operations as enabled by \proposal's bucketing (\sect{\ref{sec:mech-stage1}}), \rev{5)~\msext: a \proposal implementation
without ISP, where the same accelerators used in \proposal are  outside the SSD}, 
\rev{6)~\mscc: a \proposal configuration
in which the cores on the SSD controller\footnote{\rev{We assume three/four ARM Cortex-R4 cores~\cite{cortexr4} for \ssdc/\ssdp.}} perform \proposal's ISP tasks, and}
7)~\msfull: a \proposal~\rev{configuration in which the accelerators on the SSD controller perform the ISP tasks}. Speedup is calculated over \popt.
\aopt achieves significantly larger accuracy compared to \popt~\cite{meyer2021critical,lapierre2020metalign}. In particular, \aopt leads to 4.84$\times$ higher F1 scores and 13\% lower L1 norm scores on average across our inputs.
\unfinished{\msfull matches the accuracy of \aopt since it analyzes the same k-mer and sketch databases}.

We make six key observations. 
First, \proposal achieves significant speedup: 
on the system with \ssdc (\ssdp), \msfull is 5.34--6.38$\times$ (2.66--6.52$\times$) faster compared to \popt, and 12.41--18.22$\times$ (6.93--20.39$\times$) faster compared to \aopt.
Second, {\aopt}+\cmashopt improves \aopt's performance by 1.41$\times$ (4.19$\times$) on average, while it still performs 10.51$\times$ (2.89$\times$) slower than \msfull on \ssdc (\ssdp). This shows that while our \cmashopt approach\gram{,} even outside the SSD\gram{,} provides large benefits (specifically on \ssdp), \proposal provides more significant additional benefits by alleviating the I/O overhead. 
Third, with \ssdc (\ssdp), \msfull leads to 23.45\% (34.94\%) \hm{greater} average speedup compared to \msnol, which shows the benefits of {\proposal}'s bucketing to enable overlapping of the steps.
\rev{Fourth \msfull leads to 10.19$\times$ (2.17$\times$)
average speedup compared to \msext on \ssdc (\ssdp) due to our specialized ISP}.
\rev{Fifth,
while \mscc provides large speedups,
\msfull leads to 9\% (43\%) greater average speedup compared to \mscc on \ssdc (\ssdp). While both \proposal configurations provide large speedups, this observation shows the benefits of accelerators for \proposal's ISP tasks as the internal bandwidth grows.}
\rev{Sixth}, \proposal's speedup increases for inputs with higher genetic diversity since they lead to more sketch tree lookups in the baseline, while in \proposal's \cmashopt approach, all tax IDs are retrieved with a single pass through the tables.

\begin{figure}[h]
\centering
 \includegraphics[width=\linewidth]{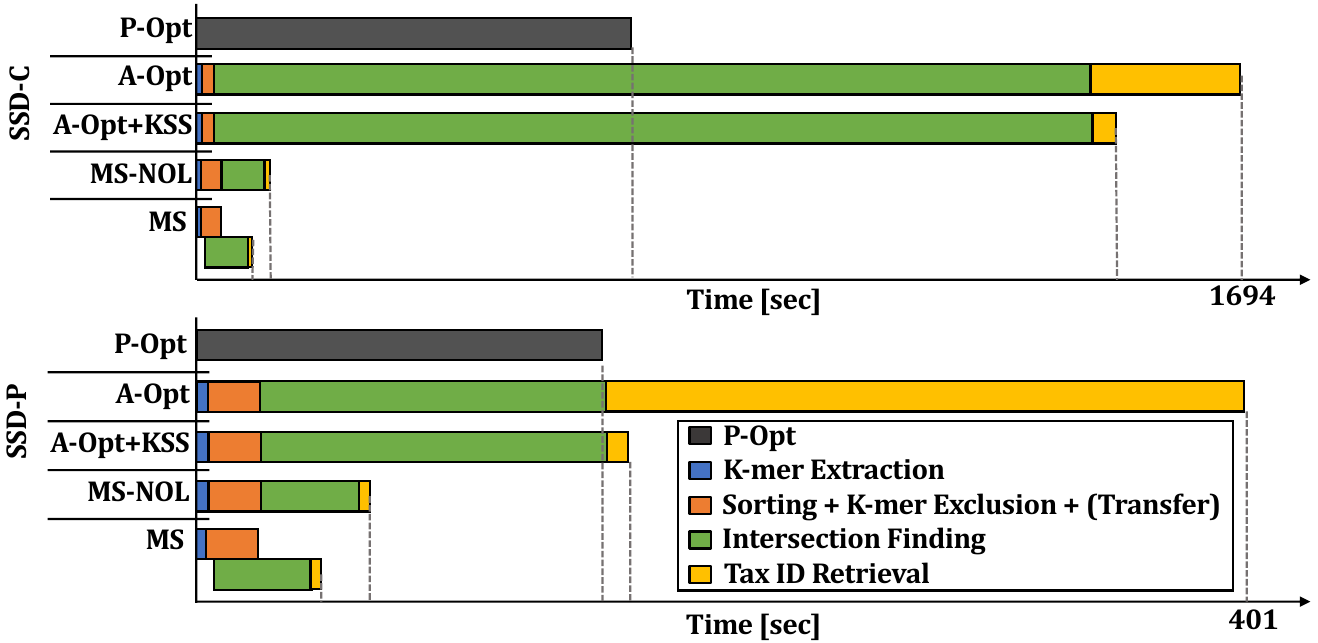}
\caption{\rev{Time breakdown with different SSDs for CAMI-L.}}
\label{fig:exec-breakdown}
\end{figure}

To further demonstrate the benefits of \proposal's optimizations, \fig{\ref{fig:exec-breakdown}}
shows the time breakdowns with \mbox{CAMI-L} as a representative. 
First, {\aopt}+\cmashopt benefits from faster tax ID retrieval compared to \aopt, 
since \cmashopt avoids many tree lookups (here 60) for each intersecting k-mer. 
Therefore, for a sufficiently large number of intersecting k-mers, \cmashopt' one-time sequential scan through all sketches is more efficient than 60 pointer-chases for \emph{each k-mer}.
Second, \msnol benefits from faster intersection finding and tax ID retrieval.
Third, \msfull benefits from overlapping part of the input query processing with intersection finding.

\begin{figure}[h]
\centering
 \includegraphics[width=\linewidth]{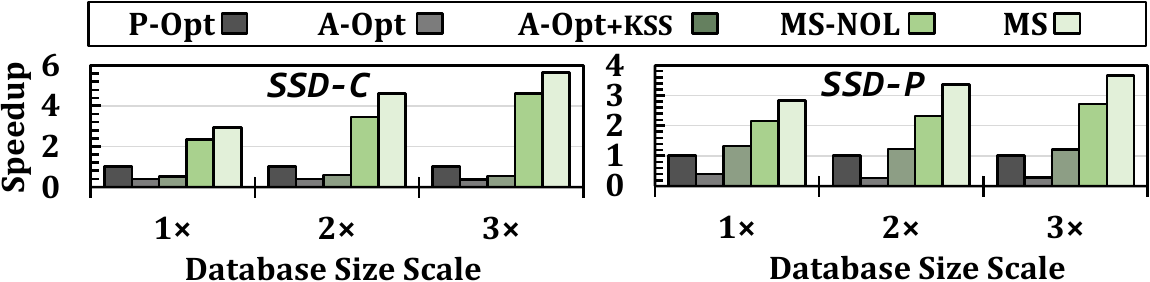}
\caption{\hm{Speedup with} different database sizes.}
\label{fig:eval-db}
\end{figure}

\head{Impact of Database Size} \fig{\ref{fig:eval-db}} shows the impact of database size, using CAMI-M as a representative input. The largest database size in each tool (marked by 3$\times$) equals the size mentioned in \sect{\ref{sec:methodology}}.
We observe that \proposal's performance benefits increase as the database size increases (up to 5.64$\times$/3.67$\times$ speedup compared to \popt 
on \ssdc/\ssdp).

\begin{figure}[h]
    \centering
    \includegraphics[width=\linewidth]{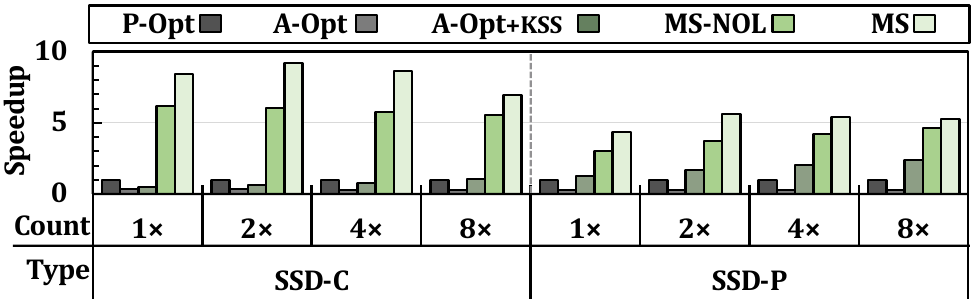}
    \caption{Speedup with different number of SSDs.}
    \label{fig:eval-ssd}
\end{figure}

\head{Impact of the Number of SSDs}
\proposal benefits from more SSDs in two ways.
First, we can map different databases to different SSDs and run various analyses concurrently, each benefiting from \proposal, as already shown. Second, since the databases and the queries are sorted in \proposal, we can disjointly partition the database between different SSDs. \fig{\ref{fig:eval-ssd}} shows \proposal's benefits in this case.
We show that \proposal maintains its large speedups with many SSDs because as the external bandwidth increases for the baselines, internal bandwidth also increases for \proposal. Particularly, speedup over \popt increases until some point (two SSDs) because \proposal takes better advantage of the bandwidth scaling due to its more efficient streaming accesses. 
Although there is a slight decrease in speedup when moving from two to eight SSDs, the speedup is still high (6.9$\times$/5.2$\times$ over eight SSD-Cs/SSD-Ps). This decrease is because in \proposal, due to the large internal bandwidth with 8 SSDs, the pipeline’s throughput becomes dependent on the sorting throughput in the host. Therefore, in large systems with many SSDs, \proposal can flexibly integrate with a sorting accelerator to further improve end-to-end performance. Based on these observations, we conclude that \proposal effectively leverages the increased internal bandwidth as the number of SSDs increases.

\begin{figure}[h]
\centering
 \includegraphics[width=\linewidth]{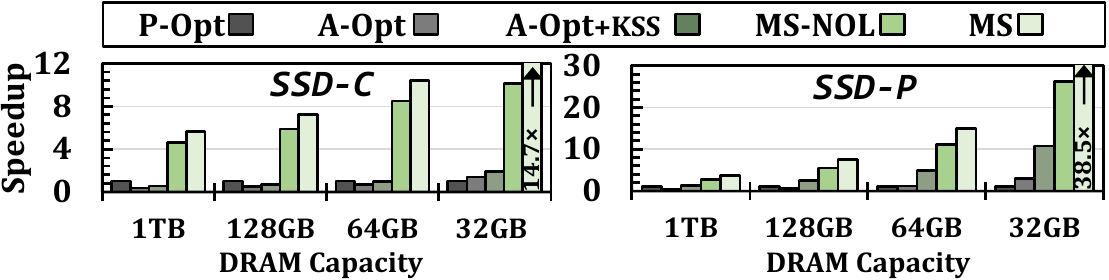}
\caption{\hm{Speedup} with different main memory capacities.}
\label{fig:eval-mem}
\end{figure}

\head{Impact of Main Memory Capacity} \fig{\ref{fig:eval-mem}} shows the impact of main memory capacity
with CAMI-M as a representative input.\footnote{In all cases, \rev{except for the 32GB configuration,} all k-mer buckets extracted from the read set (\sect{\ref{sec:mech-stage1-kmer-extraction}}) fit in DRAM.}
To gain a fair understanding of I/O overheads when DRAM is smaller than the database, we ensure to reduce this overhead as much as possible in software. To this end, we adopt an optimization~\cite{pockrandt2022metagenomic} to load and process \popt's database into chunks that fit in DRAM. In this case, random accesses to the database in each chunk do not need to repeatedly access the SSD, but there are still two overheads. First, we must still pay the I/O cost of bringing all the chunks from the SSD to DRAM. Second, for every chunk, all of the input sequences must be queried repeatedly.\footnote{Note that this optimization is not required for \aopt due to its streaming accesses to the database.}

We make three observations. 
First, \proposal's speedup increases compared to \popt with smaller DRAM (e.g., up to \rev{38.53}$\times$ on \ssdp). This is because \popt's performance gets hindered by the memory capacity, while \proposal does not rely on large DRAM for buffering the database. Second, \aopt and {\aopt}+\cmashopt  do not get affected by the small DRAM (except for the 32-GB configuration) due to their streaming database accesses with \emph{no} reuse.
But regardless of the DRAM size, they suffer from I/O overhead.
Third, with the 32-GB DRAM (which is smaller than the extracted k-mers from the read set in Step 1, \sect{\ref{sec:mech-stage1-kmer-extraction}}),
\msfull's speedup increases due to the optimizations in \proposal's bucketing (\sect{\ref{sec:mech-stage1-kmer-extraction}}).
We conclude that \proposal enables fast and accurate analysis, \emph{without relying} on large DRAM or high-bandwidth interconnects between the host and the SSD. This can 1) allow multiple jobs to run concurrently on a system and more effectively share the available DRAM capacity and SSD external bandwidth, or 2)~facilitate the use of cost-effective, portable (e.g., edge) devices for metagenomic tasks.

\newcommand\poptp{\textsf{P-Opt\_P}\xspace}
\newcommand\aoptp{\textsf{A-Opt\_P}\xspace}
\newcommand\msc{\textsf{MS\_C}\xspace}

\begin{figure}[b]
\centering
 \includegraphics[width=\linewidth]{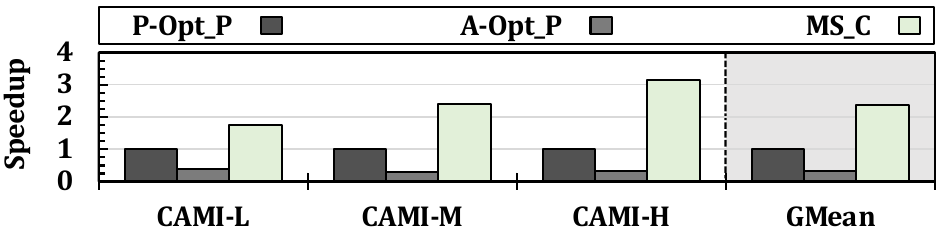}
\caption{Speedup of \proposal on a cost-optimized system over baselines on a performance-optimized system.}
\label{fig:ms-ce}
\end{figure}

\head{Impact on System Cost Efficiency} 
\proposal increases system cost-efficiency because \inum{i}~it analyzes large amounts of data inside the storage system and removes a large part of the analysis burden from other parts of the system, and \inum{ii}~does not rely on either high-bandwidth interfaces between the SSD the host or large DRAM capacity. To demonstrate this point, \fig{\ref{fig:ms-ce}}  shows \proposal's performance on a cost-optimized system with \ssdc and 64-GB DRAM (\msc) compared to \popt and \aopt on a performance-optimized system with  \ssdp and 1-TB DRAM (\poptp and \aoptp). We observe that \msc provides 2.37$\times$ and 7.20$\times$ average speedup compared to \popt and \aopt, respectively. Note that \msc provides the same accuracy as \aoptp and significantly higher accuracy compared to \poptp. We conclude that \proposal can outperform the baseline tools even when running on a more cost-optimized system and achieving high accuracy. This is critical in both increasing the overall cost-efficiency of the system and in enabling portable metagenomic analysis, which is gaining increasing importance after the advancement of compact portable DNA sequencers~\cite{minion21,jain2016oxford,cali2017nanopore} for on-site metagenomic analysis~\cite{pomerantz2018real, chiang2019from}.

\begin{figure}[h]
\centering
 \includegraphics[width=\linewidth]{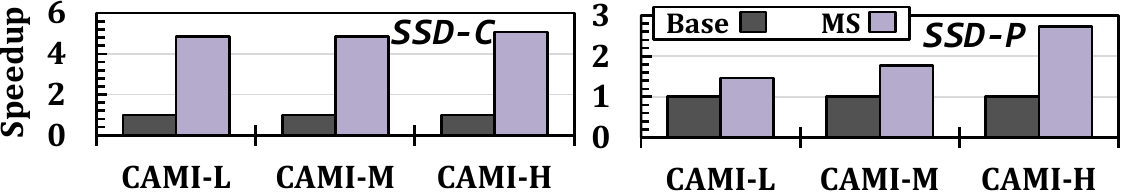}
\caption{\hm{Speedup over} a PIM-accelerated baseline~\cite{wu2021sieve}.}
\label{fig:eval-pim}
\end{figure}

\head{Comparison to a PIM Accelerator} \fig{\ref{fig:eval-pim}} evaluates \proposal compared to a PIM-accelerated baseline. We model Kraken2's \emph{end-to-end} execution time (i.e., including the I/O accesses to load data to the PIM accelerator, k-mer matching, sample classification, and other computation~\cite{wood2019improved}), \hm{performing} the k-mer matching part \hm{on} a state-of-the-art
PIM system, Sieve~\cite{wu2021sieve}.\footnote{We do not use PIM for k-mer matching in Metalign because it is bottlenecked only by the I/O bandwidth, not main memory, due to its streaming accesses.} 
We observe that \proposal achieves 4.84-5.06$\times$ (1.46-2.73$\times$) speedup on \ssdc (\ssdp) while achieving significantly higher accuracy. 
A larger database for Kraken2 to encode richer information can increase the accuracy
but will increase I/O overhead even further.
We conclude that \proposal can provide large speedups at high accuracy via analysis of large databases inside the SSD. 

\vspace{-0.5em}
\subsection{Abundance Estimation Analysis}
\label{sec:eval-abundance}

\newcommand\msnidx{\textsf{MS-NIdx}\xspace}

\begin{figure}[h]
\centering
 \includegraphics[width=\linewidth]{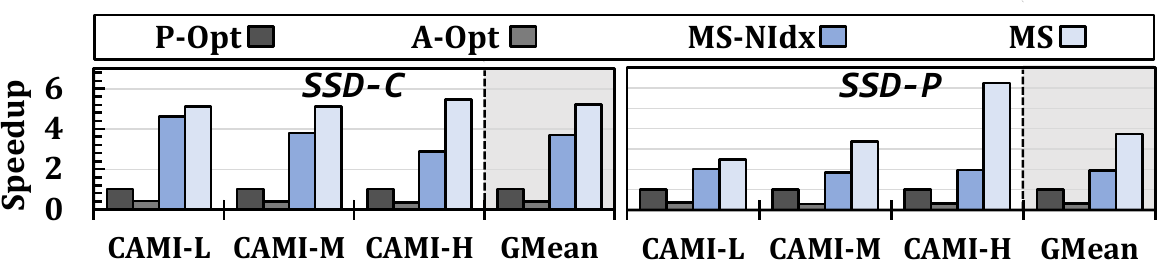}
\caption{\hm{Speedup} for abundance estimation.}
\label{fig:eval-abundance}
\end{figure}

\new{We analyze \popt, \aopt, \msnidx (a \proposal implementation \hm{that does not} leverag\hm{e} {\proposal}'s third step for generating a unified reference index shown in \sect{\ref{sec:mech-stage1}}, and instead use\hm{s} Minimap2~\cite{li2018minimap2}), and \msfull}.
\rev{In }\hm{\fig{\ref{fig:eval-abundance}}},
we make two key observations. 
First, 
\proposal leads to significant speedup compared to  \popt and \aopt, achieving 5.09--5.47$\times$ (2.46--3.72$\times$) speedup on \ssdc (\ssdp) compared to \popt, and 11.96--15.25$\times$ (6.50--20.79$\times$) speedup on \ssdc (\ssdp) compared to \popt. Second, \proposal's full implementation achieves 65\% higher average speedup compared to \msnidx.

\new{\subsection{Multi-Sample Analysis}
\label{sec:eval-multi-sample}

\begin{figure}[b]
\centering
 \includegraphics[width=\linewidth]{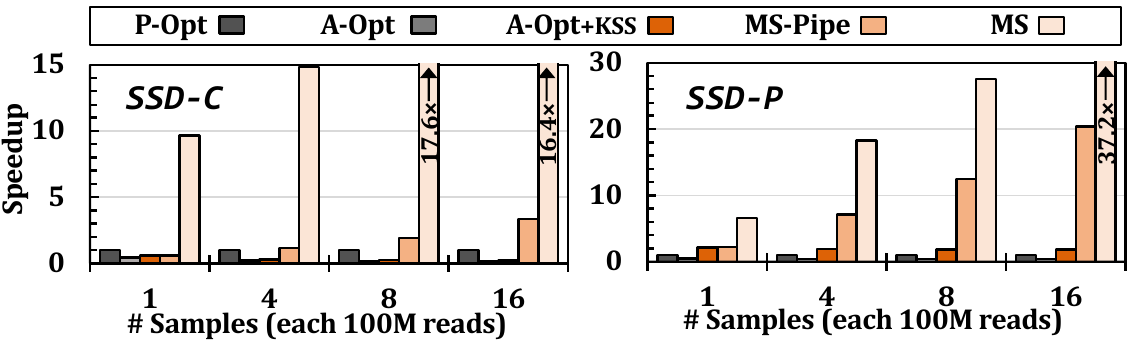}
\caption{\new{Speedup for multi-sample analysis.}}
\label{fig:multi-sample}
\end{figure}

\newcommand\mspipe{\textsf{MS-Pipe}\xspace}

\fig{\ref{fig:multi-sample}} shows speedups for multi-sample use cases (\sect{\ref{sec:mech-multi-sample}}) with 256-GB host DRAM, in which we can buffer k-mers of 1--16 samples (assuming each sample's size is similar to CAMI-M). We show the performance of our multi-sample pipelined optimization (\sect{\ref{sec:mech-multi-sample}})  in software (\mspipe) and in the full \proposal design (\msfull). In all configurations that require sorting (all except P-Opt), we use a state-of-the-art sorting accelerator~\cite{qiao2022topsort}.\footnote{\new{
We use the sorting throughput reported by the original paper~\cite{qiao2022topsort} and model the buckets' communication time with the accelerator.}} We make two key observations. First, \msfull achieves large speedups of up to 37.2$\times$/100.2$\times$ compared to \popt/\aopt. Second, \mspipe leads to up to 20.51$\times$ (52.0$\times$) speedup compared to \aopt on \ssdc (\ssdp), and the speedup grows with the number of samples. 
}

\subsection{Area and Power}
\label{sec:eval-area}

\joel{Table~\ref{tab:metastore_area_energy} shows the area and power consumption of \proposal's logic units \new{at 300 MHz}.
While they could be designed to operate at a higher frequency, their throughput is already sufficient since \proposal is bottlenecked by NAND flash reads}.
\joel{\proposal's hardware area and power requirements are minimal: only 0.04~mm$^2$ and 7.658~mW at 65~nm.
The area overhead of \proposal hardware is  0.011 mm$^2$ at 32 nm,
\footnote{We scale the area to lower technology node using the methodology in~\cite{stillmaker2017Scaling} since these technology nodes are not available publicly.}
which is 1.7\% of the three 28-nm ARM Cortex R4 cores~\cite{cortexr4} in a SATA SSD controller~\cite{samsung860pro}. \rev{These 
logic units are 26.85$\times$ more power-efficient than these cores during \proposal's ISP}.}

\begin{table}[h]
\centering
\caption{Area and power consumption of \proposal's logic.}
\label{tab:metastore_area_energy}
\resizebox{\columnwidth}{!}{%
\begin{tabular}{c|c|c|c}
\toprule
\textbf{Logic unit}                  & \textbf{\# of instances} & \textbf{Area [mm\textsuperscript{2}]} & \textbf{Power [mW]} \\ 
\midrule
\midrule
Comparator (120-bit)                 & 1 per channel              &    0.001361   &      0.284    \\
60-mer Registers (2$\times$ 120-bit)  & 1 per channel              &    0.002821   &      0.645    \\
Index Generator (64-bit)             & 1 per channel              &    0.000272   &      0.025    \\
Control Unit                         & 1 per SSD                  &    0.000188   &      0.026    \\\midrule
\textbf{Total for an 8-channel SSD}           & -                        & \textbf{0.04}    & \textbf{7.658} \\ \bottomrule
\end{tabular}
}
\end{table}

\subsection{Energy}
\label{sec:eval-energy}

For each tool, we calculate each part's energy based on its active/idle power and execution time. 
We observe that \proposal provides significant energy benefits by alleviating the I/O overhead and \new{
reducing the burdens of the analysis from the rest of the system}.
\rev{A}cross the SSDs and the datasets, we observe that \proposal leads to 5.36$\times$ (9.82$\times$), 15.16$\times$ (25.68$\times$), \hm{and} 1.88$\times$ (3.49$\times$) average (maximum) energy reduction compared to \popt, \aopt, and the PIM-accelerated \popt when finding \hm{species} present in the sample.

\section{Related Work}
\label{sec:related}

To our knowledge, \proposal is the \emph{first} ISP system for end-to-end metagenomic analysis. 
We already extensively compare \proposal against Kraken2~\cite{wood2019improved}, Metalign~\cite{lapierre2020metalign}, and Sieve~\cite{wu2021sieve}. Here, we briefly discuss other related works.

\head{Software Optimization of Metagenomics}
There are various pure software tools for metagenomics. Several tools (e.g., ~\cite{wood2014kraken,lapierre2020metalign}) use comprehensive databases for high accuracy. However, these tools usually incur significant computational and I/O costs due to their database size. 
Some tools (e.g., ~\cite{wood2019improved,pockrandt2022metagenomic,muller2017metacache}) apply sampling techniques to significantly reduce database size, but at the cost of accuracy loss. 
All of these techniques feature trade-offs between accuracy and performance due to the significant I/O costs of moving large metagenomic databases.

\head{Hardware Acceleration of Metagenomics} There are several works using GPU~\cite{jia2011metabing,kobus2021metacache,wang2023gpmeta,kobus2017accelerating}, FPGA~\cite{saavedra2020mining}, and PIM~\cite{wu2021sieve,shahroodi2022krakenonmem,shahroodi2022demeter,dashcam23micro,hanhan2022edam,zou2022biohd} to accelerate metagenomics by alleviating its computation or main memory data movement overheads. 
These works do not alleviate I/O overhead, whose impact on end-to-end performance becomes even larger when other bottlenecks get alleviated.

\head{In-Storage Processing}
Several works propose ISP designs as accelerators for different applications~\cite{liang2019cognitive,kim2020reducing,lim2021lsm,li2021glist,wang2016ssd,lee2020neuromorphic,kang2021s,han2021flash,wang2022memcore,wang2018three,han2019novel,choi2020flash,mailthody2019deepstore,pei2019registor,jun2018grafboost, do2013query, seshadri2014willow,kim2016storage, riedel2001active,riedel1998active,lee2022smartsage,jeong2019react, jun2016storage,li2023ecssd} (e.g., in machine learning~\cite{li2023ecssd,liang2019ins,lee2022smartsage}, pattern processing and read mapping~\cite{jun2016storage,mansouri2022genstore}, and graph analytics~\cite{jun2018grafboost}), general-purpose~\cite{gu2016biscuit, kang2013enabling, wang2019project,acharya1998active,keeton1998case,riedel1998active,riedel2001active,merrikh2017high,tiwari2013active,tiwari2012reducing,boboila2012active,bae2013intelligent,torabzadehkashi2018compstor,kang2021iceclave}, bulk-bitwise operations using flash memory~\cite{gao2021parabit,park2022flash}, in close integration with FPGAs~\cite{jun2015bluedbm, jun2016bluedbm, torabzadehkashi2019catalina, lee2020smartssd, ajdari2019cidr, koo2017summarizer}, or GPUs~\cite{cho2013xsd}. 
None of these works perform metagenomic analysis nor address the challenges of ISP for metagenomics.

\proposal is the first work that targets the end-to-end process of metagenomic analysis, and by addressing the unique challenges of leveraging ISP for metagenomics, \proposal fundamentally alleviates its data movement overhead from the storage system.  \proposal does so through its lightweight, synergistic, and cooperative design with processing inside and outside the storage system.

\section{Conclusion}
\label{sec:conclusion}

We introduce \proposal, the first ISP system for metagenomics to reduce the I/O cost of metagenomics. To enable efficient ISP for metagenomics, we propose new 1) task partitioning, 2) storage technology-aware algorithms, 3) data mapping, and 4) data/computation flow coordination. Our evaluations on large metagenomic databases show that \proposal improves \gram{the} performance and energy efficiency of metagenomics, with high accuracy and at \gram{a} low cost.

\bibliographystyle{IEEEtran}
\bibliography{refs}

\end{document}